\documentclass[aps,prb,twocolumn,superscriptaddress,preprintnumbers]{revtex4}

\usepackage[colorlinks,bookmarks=false,citecolor=blue,linkcolor=red,urlcolor=blue]{hyperref}
\usepackage{epsfig}
\usepackage{tikz}
\usepackage{graphicx}
\usepackage{physics}

\usepackage{dcolumn}
\usepackage{bm}

\usepackage{amsmath,amssymb}

\makeatletter
\renewcommand*\env@matrix[1][*\c@MaxMatrixCols c]{%
  \hskip -\arraycolsep
  \let\@ifnextchar\new@ifnextchar
  \array{#1}}
\makeatother

\usepackage{appendix}

\begin{document}

\title{The quantum spin quadrumer}

\author{Subhankar Khatua}
\email{subhankark@imsc.res.in}
\affiliation{The Institute of Mathematical Sciences, C I T Campus, Chennai 600 113, India}
\affiliation{Homi Bhabha National Institute, Training School Complex, Anushakti Nagar, Mumbai 400094, India}
\author{R. Shankar}
\email{shankar@imsc.res.in}
\affiliation{The Institute of Mathematical Sciences, C I T Campus, Chennai 600 113, India}
\affiliation{Homi Bhabha National Institute, Training School Complex, Anushakti Nagar, Mumbai 400094, India}
\author{R. Ganesh}
\email{ganesh@imsc.res.in}
\affiliation{The Institute of Mathematical Sciences, C I T Campus, Chennai 600 113, India}
\affiliation{Homi Bhabha National Institute, Training School Complex, Anushakti Nagar, Mumbai 400094, India}

\date{\today}

\begin{abstract}
A fundamental motif in frustrated magnetism is the fully mutually coupled cluster of $N$ spins, with each spin coupled to every other spin. Clusters with $N=2$ and $3$ have been extensively studied as building blocks of square and triangular lattice antiferromagnets. 
In both cases, large-$S$ semiclassical descriptions have been fruitfully constructed, providing insights into the physics of macroscopic magnetic systems. 
Here, we develop a semiclassical theory for the $N=4$ cluster. This problem has rich mathematical structure with a ground state space that has non-trivial topology. 
We show that ground states are appropriately parametrized by a unit vector order parameter and a rotation matrix. Remarkably, in the low energy description, the physics of the cluster reduces to that of an emergent free spin-$S$ spin and a rigid rotor. This 
successfully explains the spectrum of the quadrumer and its associated degeneracies.  
However, this mapping does not hold in the vicinity of collinear ground states due to a subtle effect that arises from the non-manifold nature of the ground state space. We demonstrate this by an analysis of soft fluctuations, showing that collinear states have a larger number of soft modes. 
Nevertheless, as these singularities only occur on a subset of measure zero, the mapping to a spin and a rotor provides a good description of the quadrumer. We interpret thermodynamic properties of the quadrumer that are accessible in molecular magnets, in terms of the rotor and spin degrees of freedom.
Our study paves the way for field theoretic descriptions of systems such as pyrochlore magnets.
\end{abstract}
\keywords{}
\maketitle

\section{Introduction:}
The principles underlying frustrated magnetism emerge from a few prototypical models. Many of these share a common feature: they are composed of clusters of $N$ spins with each spin equally coupled to every other spin. Such a cluster is described by the Hamiltonian 
\begin{equation}
 H_N = J \left\{\sum_{j=1}^N \vec{S}_j \right\}^2.
\end{equation}
Frustration emerges when $J>0$, representing antiferromagnetic coupling between each pair of spins. When the clusters are coupled among themselves, this typically leads to effects such as macroscopic classical degeneracy. 
For example, clusters of $N=2$ spins occur in the square antiferromagnet and in dimerized quantum systems such as SrCu$_2$(BO$_3$)$_2$\cite{Shastry1981,Kageyama1999}. Clusters with $N=3$ occur in the Majumdar Ghosh model\cite{MajumdarGhosh}, the triangular antiferromagnet and the kagome antiferromagnet. 

For systems with $N=2$ and $N=3$ clusters, a particularly fruitful approach has been to construct large-$S$ semiclassical field theories. The field theory for $N=2$ systems, first derived by Haldane\cite{Haldane1983}, is formulated in terms of a unit vector field. On the other hand, the $N=3$ field theory is more appropriately written in terms of an $SO(3)$ rotor field as first shown by Dombre and Read\cite{Dombre1989}. 
A similar field theoretic approach has so far not been realized for $N=4$. This is an interesting and topical problem due to its relevance to pyrochlore antiferromagnets\cite{Gardner2010}, the checkerboard lattice antiferromagnet\cite{Canals2002} and the square J$_1$-J$_2$-J$_3$ model\cite{Danu2016}. In particular, it is relevant to several pyrochlore materials with Heisenberg-like couplings such as Mn$_2$Sb$_2$O$_7$, CdYb$_2$S$_4$, Gd$_2$Ti$_2$O$_7$, etc., which all have intriguing properties\cite{Gardner2010}. 

Here, we derive a path integral description for the $N=4$ cluster which serves as a starting point for constructing semiclassical field theories. Even at the level of a single cluster, we find rich topological structure and an elegant physical description.

\section{Cluster ground states for $N=2,3$}
Classically, a spin is a vector of length $S$. The allowed values of spin form a one-to-one and onto mapping to $S^2$, the two-dimensional sphere. An arbitrary spin position can be described by two parameters, e.g., polar and azimuthal angles. 
As the Hamiltonian $H_N$ is positive semi-definite, the lowest possible classical energy is zero. In other words, a ground state is reached when the total spin is zero, i.e., $\sum_{j=1}^N \vec{S}_j =0$. The set of all such $N$-spin states constitutes the ground state space. 
Mathematically, this can be denoted as $\{ S^2 \otimes S^2 \otimes \ldots \otimes S^2 \vert \vec{S}_1 + \ldots +\vec{S}_N =0\}$. 
The special features of the $N=4$ cluster lie in the topology of its classical ground state space. We first recapitulate the properties of the $N=2$ and $N=3$ clusters to set the stage for $N=4$. 

For $N=2$, the ground state space is simply the set of pairs of antipodal points on the sphere. Each ground state is uniquely defined by the position of the first spin, with the ground state space being isomorphic to $S^2$. This mapping brings out the topology of the ground state space, e.g., showing that it is simply connected. It also brings out its `manifold' character as every point in $S^2$ has a two-dimensional tangent space. In physics terms, about any given ground state, we have two independent `soft' fluctuations that do not cost energy. This mapping to $S^2$ underlies the semi-classical field theory for the antiferromagnetic Heisenberg chain. First formulated by Haldane\cite{Haldane1983}, the field theory is written in terms of a slowly-varying field, $\hat{n}(x,t) \in S^2$. 

For $N=3$, the ground states are 120$^\circ$ states -- the three spins lie in a plane forming the sides of an equilateral triangle. All such states can be obtained from a global rotation operation performed on a reference 120$^\circ$ state, say in the $XY$ plane. Thus, each ground state can be uniquely mapped to an $SO(3)$ rotation matrix. The ground state space is thus isomorphic to $SO(3)$. As before, this forms a manifold, i.e., at any point in $SO(3)$, there exists a three dimensional tangent space. Every ground state allows for three independent soft fluctuations. Naturally, a semi-classical field theory for $N=3$ systems, e.g., the triangular lattice antiferromagnet and the Majumdar Ghosh model, is formulated in terms of a matrix field, $R (x,t) \in SO(3)$\cite{Dombre1989,RAO1994547}.

\section{Parametrizing the $N=4$ classical ground state space}

The $N=4$ case presents a non-trivial step forward from the $N=2$ and $N=3$ cases. We first enumerate the degrees of freedom. The total space is eight dimensional ($S^2 \otimes S^2\otimes S^2\otimes S^2$) as each spin has two independent parameters. The constraint of zero total spin, $\sum_{j=1}^N \vec{S}_j =0$, is, in fact, three independent contraints -- one for each component of the total spin. With eight degrees of freedom and three constraints, the ground state space is five-dimensional. Naively, we may expect the set of ground states to form a five-dimensional manifold. However, we show below that a much more nuanced picture emerges. 

Several parametrizations of the ground state space, with minor variations, are available in literature\cite{Reimers1991,Moessner1998b, Tchernyshyov2002,Chalker2011,Plat2015,Danu2016,Wan2016}.  Here, we present a parametrization that leads to an elegant semi-classical description. 
\begin{figure}
\includegraphics[height=44mm, width=40mm]{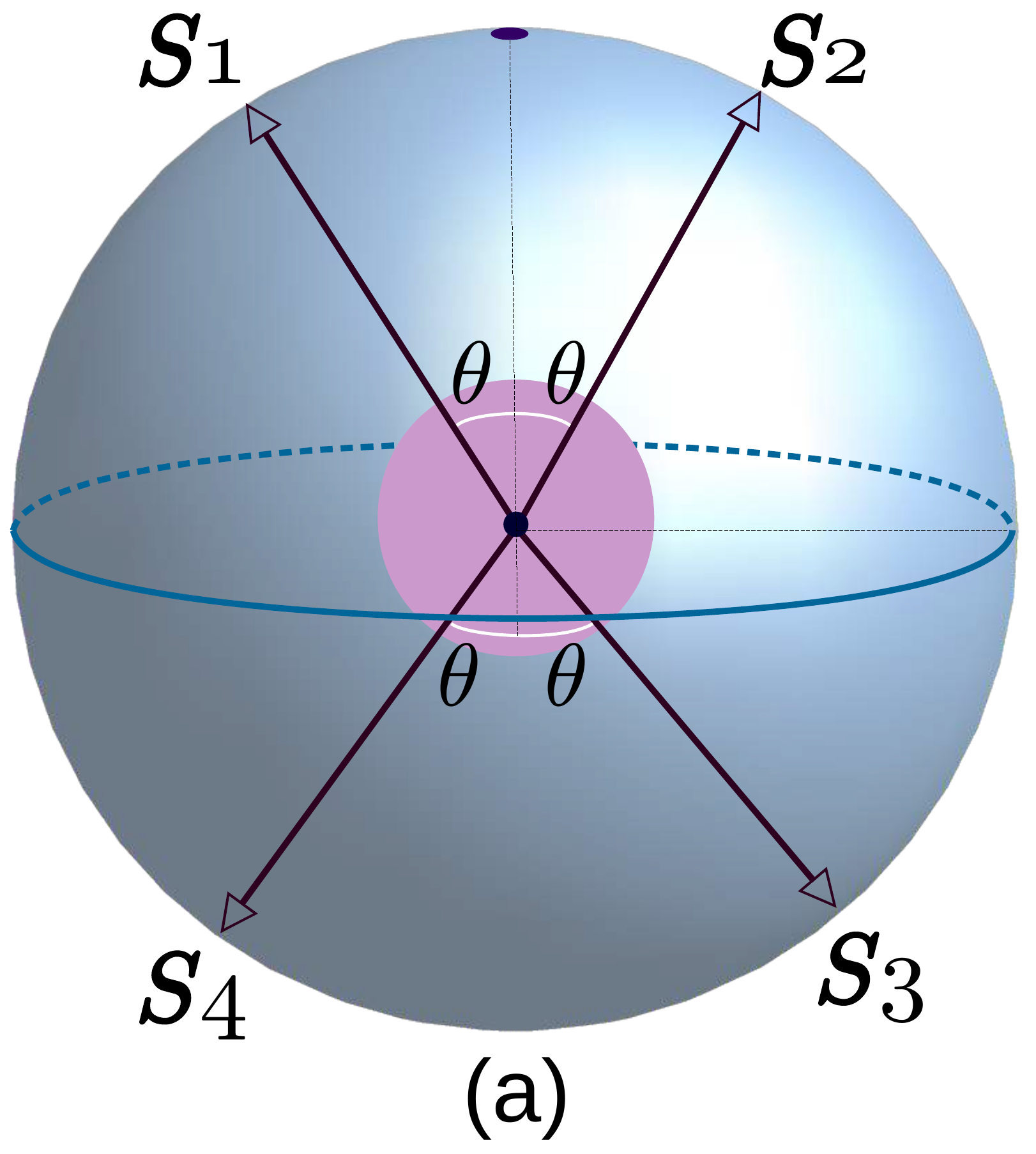}
\includegraphics[height=45mm, width=41mm]{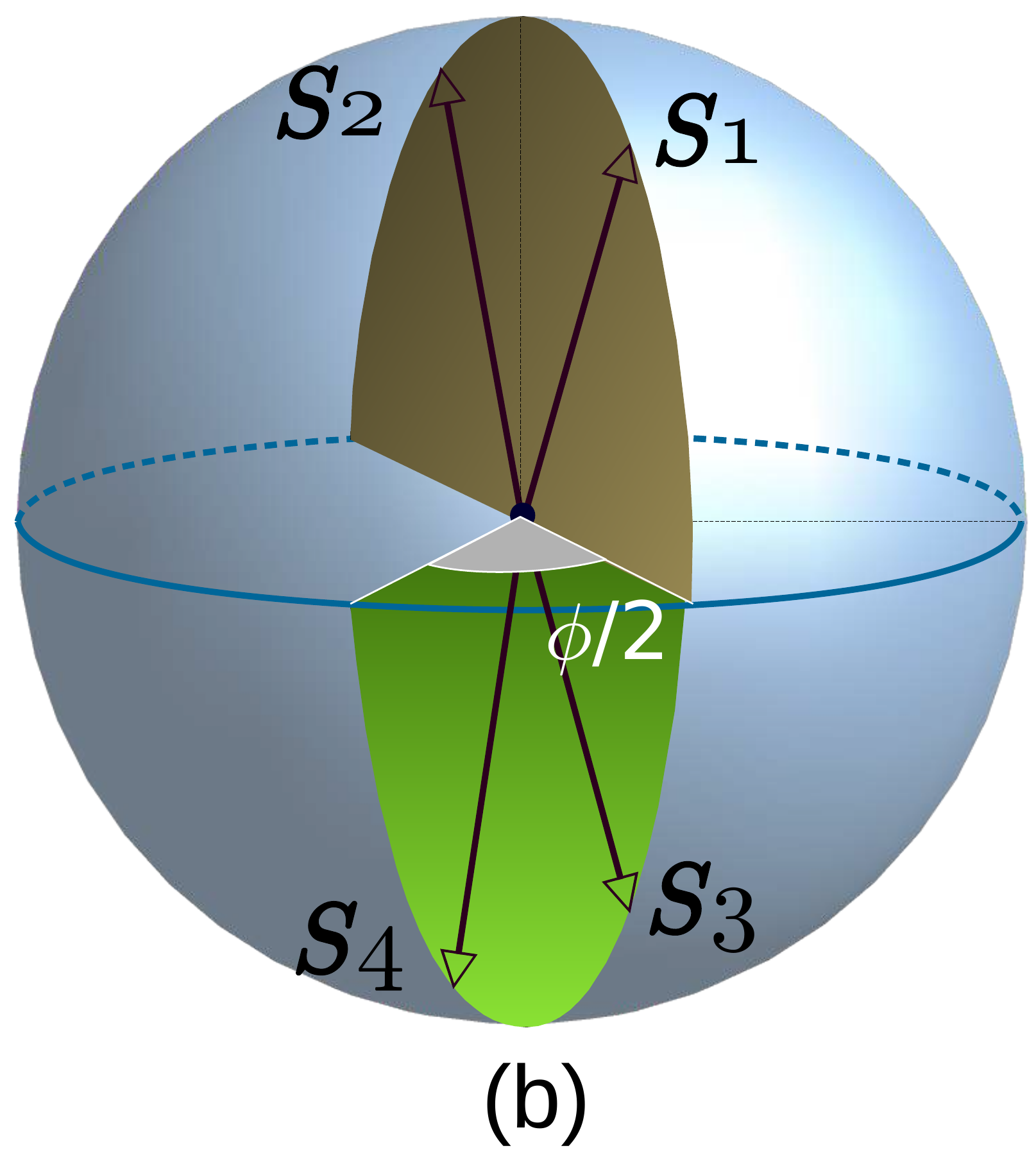}
\caption{Parametrizing the ground state space: (a) We initially take all spins to lie in the $XZ$ plane with $\vec{S}_1 = -\vec{S}_3$ and $\vec{S}_2 = -\vec{S}_4$. The angle between $\vec{S}_1$ and $\vec{S}_2$ is taken to be $2\theta$. (b)  We now rotate the spins pairwise about an axis that lies along $\vec{S}_1+\vec{S}_2$. As a result, the plane containing $\vec{S}_1$ and $\vec{S}_2$ makes an angle $\phi/2$ with the plane containing $\vec{S}_3$ and $\vec{S}_4$. Upto a global rotation, all ground states are described by choosing appropriate values of $\theta$ and $\phi$. }
\label{fig1}
\end{figure}

A generic ground state can be described using five parameters, as shown in Fig.~\ref{fig1}. To construct this state, we initially choose all four spins to lie in the $XZ$ plane, with $\vec{S}_1$ and $\vec{S}_2$ subtending an angle $2\theta$ while their sum points along the $Z$ axis. This constrains $\vec{S}_3$ and $\vec{S}_4$ to also subtend the same angle $2\theta$ with their sum pointing along $-\hat{z}$. We  now introduce the second degree of freedom $\phi$; we rotate $\vec{S}_1$ and $\vec{S}_2$ by $\phi/4$ about the $Z$ axis. At the same time, we rotate $\vec{S}_3$ and $\vec{S}_4$ by $-\phi/4$. This operation preserves $\vec{S}_1 + \vec{S}_2$ and $\vec{S}_3+\vec{S}_4$. This prescription leads to four unit vectors,
\begin{eqnarray}
\hat{n}_1 &=& \sin\theta\left(\cos\frac{\phi}{4}\;\hat{x}+\sin\frac{\phi}{4}\;\hat{y}\right) + \cos\theta \;\hat{z},\nonumber\\
\hat{n}_2 &=& \sin\theta\left(-\cos\frac{\phi}{4}\;\hat{x}-\sin\frac{\phi}{4}\;\hat{y}\right) + \cos\theta \;\hat{z},\nonumber\\
\hat{n}_3 &=& \sin\theta\left(-\cos\frac{\phi}{4}\;\hat{x}+\sin\frac{\phi}{4}\;\hat{y}\right) -\cos\theta \;
\hat{z}, \nonumber\\
\hat{n}_4 &=& \sin\theta\left(\cos\frac{\phi}{4}\;\hat{x}- \sin\frac{\phi}{4}\;\hat{y}\right) - \cos\theta \;
\hat{z}.
\label{eq.nis}
\end{eqnarray}
It is easy to see that these four vectors have unit length and add to zero. This configuration describes a generic ground state, modulo a global spin rotation.

Here, we constrain $\theta \in [0,\pi)$ and $\phi \in [0,2\pi]$. These ranges 
allow for a faithful representation of allowed ground states without double counting~\footnote{
For example, a configuration with $(\theta=\alpha,\phi=2\pi+\beta)$ is equivalent to one with $(\theta = \pi-\alpha,\phi=2\pi-\beta)$ modulo a $\pi$-rotation about the Y axis.}. The parameters $\theta$ and $\phi$ describe relative angles between spins. However, as seen from their specified ranges, they resemble a unit-vector with polar angle $\theta$ and azimuthal angle $\phi$, which encodes the internal state of the cluster.

Finally, the four spins in the cluster are given by $\vec{S}_j = SR \hat{n}_j$, incorporating three degrees of freedom into $R$, an $SO(3)$ rotation matrix. It can be seen that all possible ground states are described by appropriate choices of these five parameters: $\theta$, $\phi$, and three Euler angles describing $R$. In this sense, the ground state space is indeed five-dimensional. Naively, we may guess that the space is simply $SO(3)\otimes S^2$ with $S^2$ being the space of the vector described by $\theta$ and $\phi$. However, as we show below, the ground state space has non-trivial topology with non-manifold character.

Some representative members of the ground state space are: (i) a tetrahedral state with spins pointing towards the corners of a regular tetrahedron, (ii) a coplanar state with spins forming the sides of a square, and (iii) a collinear state with $\vec{S}_1 = \vec{S}_2 = -\vec{S}_3 =- \vec{S}_4$. As discussed in Ref.~\onlinecite{Danu2016}, the ground states may be classified as non-coplanar, coplanar and collinear. In particular, all coplanar states can be accessed either by setting $\phi=0$ or by setting $\theta = \frac{\pi}{2}$ in the above parametrization. To have a collinear state, the four spins must form two pairs of parallel spins which are anti-aligned with respect to each other. This leads to three distinct collinear states (upto a global spin rotation) -- this is the number of ways of forming two pairs from four objects. The three distinct collinear states correspond to $\{\vec{S}_1=\vec{S}_2=-\vec{S}_3=-\vec{S}_4\}$, $\{\vec{S}_1=\vec{S}_3=-\vec{S}_2=-\vec{S}_4\}$ and $\{\vec{S}_1=\vec{S}_4=-\vec{S}_2=-\vec{S}_3\}$. We will see below that these collinear states play a key role in the topology of the ground state space.

\section{Path integral description}

To describe the $N=4$ cluster in the large-$S$ semi-classical limit, we develop a path integral formulation. We first parametrize the spins as 
\begin{eqnarray}\label{eq3}
 \vec{S}_j = S\hat{\Omega}_j\approx S R(\hat{n}_j+M_j\frac{\vec{L}}{S}).
\label{eq.S_parametrization}
 \end{eqnarray}
Here, $R$ represents an $SO(3)$ rotation matrix while $\hat{n}_j$'s represent unit vectors determined by $\theta$ and $\phi$, as defined in Eq.~\ref{eq.nis}. The vector $\vec{L}$ is a new parameter that encodes net magnetization. In other words, $\vec{L}$ represents the deviation from the ground state space. Note that $\vec{L}$ has three independent components. Together with $\theta$, $\phi$, and $R$, this accounts for the eight degrees of freedom that determine the space of all allowed configurations. As we are interested in a low energy effective theory, we take $\vec{L}$ to be small. We take the spin length to be large, $S\gg 1$, while assuming $\vec{L}\sim \mathcal{O}(1)$ so that the deviation from the ground state space is $\mathcal{O}(1)$. The factor of $1/S$ that comes with $\vec{L}$ serves as a convenient bookkeeping tool. Below, we derive the path integral partition function as an expansion in powers of $S$, keeping terms upto $\mathcal{O}(S^0)$ in the action and neglecting all terms with lower powers of $S$.

We have introduced a matrix, $M_j$, given by $M_j^{\alpha\beta}=\delta^{\alpha\beta}-n_j^\alpha n_j^\beta$. This matrix is, in fact, the projector onto the plane perpendicular to $\hat{n}_j$.
It guarantees that the vector $\hat{\Omega}_j$ is normalized to $\mathcal{O}(S^0)$. In all calculations below, we take $S$ to be large and keep terms to $\mathcal{O}(S^0)$ in the action. 

The  magnetization of the cluster is now given by, 
\begin{eqnarray}
\sum_{j=1}^4 \vec{S_j} =R\sum_{j=1}^4 M_j\vec{L} \equiv R\left(M\vec{L}\right),
\label{eq.netmoment}
\end{eqnarray}
where $M^{\alpha\beta}=\sum_{j=1}^4 M_j^{\alpha\beta}= 4\delta^{\alpha\beta}- \sum_j n_j^\alpha n_j^\beta $. We note here that the magnetization vector is an angular momentum variable as it is a sum of spins. Upon quantization, its components should satisfy angular momentum commutation relations.

We follow the well-established semi-classical path integral formalism for spin-$S$ spins\cite{Auerbach_book}. For our cluster of $N=4$ spins, the partition function is given by 
\begin{eqnarray}
 \mathcal{Z} &=& \int\,\mathcal{D}\hat{\Omega}_1\,\mathcal{D}\hat{\Omega}_2\,\mathcal{D}\hat{\Omega}_3\,\mathcal{D}\hat{\Omega}_4 \times \nonumber\\
&\phantom{a}&\delta (\Omega^2_1 -1)\,\delta (\Omega^2_2 -1)\delta (\Omega^2_3 -1)\delta (\Omega^2_4 -1)e^{-\mathcal{S}},
\end{eqnarray}
where $\mathcal{S}$ is the action given by,
\begin{equation}
\mathcal{S}=\int_0^\beta d\tau\left(i\sum_{j=1}^4\vec{A}(\hat{\Omega}_j)\cdot\partial_\tau \vec{S}_j +J\left(\sum_{j=1}^4 \vec{S}_j\right)^2\right).
\end{equation}
The path integral is over the three components of $\hat{\Omega}_{1,\ldots,4}$ which are integrated over the real line at every imaginary time slice. 
The $\delta$ functions in the integrand ensure normalization. 

\subsection{Berry phase term}

The first term in the action is the Berry phase with $\vec{A}$ defined as $\epsilon_{\alpha\beta\gamma}\partial_\beta A^\gamma(\hat{\Omega}) = \Omega^\alpha$. Essentially, $\vec{A}$ is the vector potential of a magnetic monopole at the origin, with total flux $4\pi$. The integral, $\int_0^\beta d\tau i\vec{A}(\hat{\Omega}_j)\cdot\partial_\tau \vec{S}_j$, is a geometric quantity, equal to $iS$ times the area covered by $\hat{\Omega}_j(\tau)$ on the surface of the unit sphere.

We evaluate the Berry phase to $\mathcal{O}(S^0)$,
\begin{eqnarray}\label{eq7}
\nonumber  i\int_0^\beta d\tau \sum_j\vec{A}(\hat{\Omega}_j)\cdot\partial_\tau \vec{S}_j &=&  
\\
\int_0^\beta \!\! d\tau \!
\Big[iS\sum_{j=1}^4 \! \vec{A}(R\hat{n}_j) \! \cdot \!\partial_\tau(R\hat{n}_j) 
\!\!\! &+& \!\!\! 4i\vec{L} \! \cdot \! \vec{U}- i\vec{V} \!\cdot \! RM\vec{L}\Big]\!,\phantom{abc}
\label{eq.berry}
\end{eqnarray}
where we have defined two vector quantities, $\vec{U}= \frac{1}{4} \sum_j \partial_\tau \hat{n}_j\times \hat{n}_j $ and $V_\beta = -\frac{1}{2}\epsilon_{\beta\sigma\delta}\{(\partial_\tau R)R^{-1}\}^{\sigma\delta}$. Repeated indices are to be summed over. The vector $V_\beta$ has an identifiable form; it is the angular velocity of a rigid body whose orientation is described by the matrix $R$. To arrive at Eq.~\ref{eq.berry}, we have used two identities: $\epsilon_{\alpha\sigma\rho} R^{\alpha\beta}  R^{\rho\delta} R^{\sigma\sigma'} = \epsilon_ {\beta\sigma'\delta}$ and $\epsilon_{\alpha\beta\gamma}\partial_\beta A^\gamma(\hat{\Omega}) = \Omega^\alpha$.

The vector $\vec{U}$ depends purely on $\hat{n}_j$'s, and thereby on $\theta$ and $\phi$. 
Remarkably, $\vec{U}$ uniformly vanishes for any choice of $\theta$ and $\phi$. This is ultimately due to the symmetric parametrization of $\hat{n}_j$'s in terms of the $\theta$ and $\phi$. Following further simplifications (see Appendix.~\ref{appA}), the Berry phase term comes out to be
\begin{eqnarray}
 \int_0^\beta d\tau\,\left(-iS \cos\theta\dot\phi - i RM\vec{L}\cdot \vec{V}\right).
\end{eqnarray}
Note that the Berry phase decouples into two terms: the first only depends on the parameters $\theta$ and $\phi$, while the second contains the $SO(3)$ matrix variable $ R$. The second term also depends on $\theta$ and $\phi$, via the matrix $M$. Remarkably, the first term is precisely the Berry phase of a spin-$S$ spin. We had earlier discussed that $\theta$ and $\phi$ variables resemble a unit vector order parameter. Here, from the form of the Berry phase term, we see that this vector, in fact, behaves as a spin-$S$ spin.

\subsection{Energy term}
The energy term in the action is simply $\int_0^\beta d\tau J(M\vec{L})^2$. The energy scales as the square of $\vec{L}$, which represents deviation from the ground state space. Notably, the Hamiltonian also depends on the ground state parameters $\theta$ and $\phi$ which determine the matrix $M$.

\subsection{Path integral measure}

The partition function, in terms of the new variables, becomes 
\begin{eqnarray}
\mathcal{Z} = \int \left\{ \prod_\tau \mathcal{J}(\theta_\tau,\phi_\tau,\alpha_\tau,\beta_\tau,\gamma_\tau,\vec{L}_\tau)\times \right.\nonumber \\
\phantom{abcd}\left.\,d\theta_\tau\,d\phi_\tau\,d\alpha_\tau\,d\beta_\tau\,d\gamma_\tau\,d\vec{L}_\tau\right\} e^{-\mathcal{S}},
\end{eqnarray}    
where the index $\tau$ denotes imaginary time slices.
The action $\mathcal{S}$ is given by 
\begin{eqnarray}
\mathcal{S}=\int_0^\beta d\tau \Big(-iS \cos\theta \dot\phi  - i RM\vec{L}\cdot \vec{V}
+ J(M\vec{L})^2\Big).\phantom{a}
\end{eqnarray}
We have introduced three angles, $\alpha$, $\beta$ and $\gamma$, to parametrize the rotation matrix $R$. The parameters $\alpha$ and $\beta$ determine an axis of rotation, while $\gamma$ specifies the angle of rotation about this axis. This parametrization leads to a convenient form for the path integral measure\cite{Jones_book}. 

The quantity $\mathcal{J}(\theta_\tau,\phi_\tau,\alpha_\tau,\beta_\tau,\gamma_\tau,\vec{L}_\tau)$ denotes the Jacobian for the transformation given by Eq.~\ref{eq.S_parametrization}. To $\mathcal{O}(S^0)$, the Jacobian for a given time slice takes the form $\mathcal{J}\propto\frac{1}{4\pi^2}\sin^2(\frac{\gamma}{2})\sin\alpha\, Det(M)\sin\theta$, see Appendix.~\ref{appB} for a detailed derivation. 
We only keep $\mathcal{O}(S^0)$ terms in the Jacobian. Higher order corrections, upon exponentiation, give rise to subleading $\mathcal{O}(S^{-1})$ terms in the action. 
These terms can be ignored as we only keep terms up to $\mathcal{O}(S^0)$ in the action. 
We find 
\begin{eqnarray}
 \nonumber \mathcal{J}(\theta_\tau,\phi_\tau,\alpha_\tau,\beta_\tau,\gamma_\tau,\vec{L}_\tau) \approx \phantom{abcdefgh}\\
 \frac{1}{4\pi^2}\sin^2(\gamma_\tau/2)\sin\alpha_\tau \sin\theta_\tau Det(M).
 \label{eq.Jacobian}
\end{eqnarray}
This form has an elegant interpretation as a measure for the path integral. It contains the $SO(3)$ group-invariant measure\cite{Jones_book}: $dR \sim \frac{1}{4\pi^2}\sin^2(\gamma/2)\sin\alpha\,d\alpha\,d\beta\,d\gamma$. We also identify a measure for the emergent vector defined by $\theta$ and $\phi$: $d\hat{\Omega}_{\theta,\phi} = \sin\theta\,d\theta\,d\phi$.
The factor $Det(M)$ can be absorbed into the infinitesimal $d\vec{L}$ by redefining 
$\vec{L}' = R(M\vec{L})$, which is the net moment of the cluster defined in Eq.~\ref{eq.netmoment}. 
Note that $M$ is a $3\times3$ matrix that depends on $\theta$ and $\phi$. As $ R$ is an $SO(3)$ rotation matrix, its determinant is unity. 

The partition function becomes 
\begin{eqnarray}
\nonumber \mathcal{Z} &=& \int \left\{\prod_\tau d\hat{\Omega}_{\theta_\tau,\phi_\tau} \,d \vec{L}'_\tau\,d{R}_\tau\right\} e^{-\mathcal{S}} \\
&=& \int \mathcal{D}\hat{\Omega}_{\theta,\phi} \,\mathcal{D} \vec{L}'\,\mathcal{D}{R}\, e^{-\mathcal{S}} .
\end{eqnarray}
Note that we have implicitly assumed an order of integration, viz., that $\vec{L}'$ will be integrated out before $\hat{\Omega}_{\theta,\phi}$ variables. This is necessary as the definition of $\vec{L}'$ involves the matrix $M$ which depends on $\theta$ and $\phi$ -- we will see below that this dependence brings out the non-manifold character of the ground state space. 

The path integral action is given by $\mathcal{S}=\int_0^\beta d\tau\,\left(-iS \cos\theta\dot\phi - i\vec{L}'\cdot \vec{V}+ JL'^2\right)$, where the vector $\vec{V}$ is defined above. Remarkably, with our choice of order of integration, the partition function apparently decouples into two parts,
\begin{eqnarray}
\mathcal{Z} &=& \left(\int \mathcal{D}\hat{\Omega}(\theta,\phi)\,e^{\int_0^\beta d\tau\,iS \cos\theta\dot\phi}\right)\nonumber\\
&\times&\left(\int\mathcal{D}\vec{L}'\,\mathcal{D} R\, e^{-\int_0^\beta d\tau \left(-i\vec{L}'\cdot \vec{V} +JL'^2\right)}\right)\nonumber\\
&\equiv&\mathcal{Z}_1 \times \mathcal{Z}_2.
\label{eq.partitionfunction}
\end{eqnarray}
Both $\mathcal{Z}_1$ and $\mathcal{Z}_2$ are well known paradigmatic forms. $\mathcal{Z}_1$ is the partition function of a free spin-$S$ moment. 
This spin is `emergent' -- it is not a microscopic variable, but an encoding of the internal degrees of freedom, $\theta$ and $\phi$. 
$\mathcal{Z}_2$ represents the partition function of a spherical top (a rigid rotor with the three principal moments equal) with moment of inertia $\frac{1}{2J}$. 
The matrix $R$ represents angular position, while $\vec{L}'$ represents angular momentum. 
Note that $\vec{L}'$ is the total moment of the cluster, with its components obeying angular momentum commutation relations. It represents `hard modes' that can be integrated out to obtain a zero-temperature description.

This is the main result of this article: the system of four spins coupled by mutual antiferromagnetic interactions, in the semi-classical low-energy limit, decouples into a rigid rotor and an emergent free spin-$S$ spin!

\section{Comparison with conventional quantum analysis}

To check for consistency of the mapping to a spin and a rotor, we compare the energy spectrum given by this mapping to that obtained from a conventional quantum analysis. 
Conventionally, finding the spectrum of the Hamiltonian $H_4 = J(\vec{S}_1 + \vec{S}_2 + \vec{S}_3 + \vec{S}_4)^2$ reduces to a problem of angular momentum addition. The energy eigenvalues are simply $J j(j+1)\hbar^2$ with $j=0,\dots,4S$ being the total spin quantum number. 

In the semi-classical approach, we have a free spin-$S$ moment and a rigid rotor. 
The free spin does not contribute to energy as its Hamiltonian is zero. The rigid rotor with moment of inertia $\frac{1}{2J}$ does contribute, with the spectrum known to be precisely $J j(j+1)\hbar^2$ with $j=0,\dots,\infty$\cite{casimir}. Thus, we obtain the same low-energy spectrum from the semi-classical as well as the fully quantum approach. 

To further characterize the spectrum, we obtain the degeneracy of each level using both approaches. The obtained degeneracies are compared in Table.~\ref{tab.degen}.
The calculation of degeneracy using the conventional `full quantum' approach is discussed in Appendix.~\ref{appC}. As for the semi-classical approach, the degeneracy of the rigid rotor problem\cite{casimir} is well known to be $(2j+1)^2$. This is to be multiplied by $(2S+1)$ on account of the free spin. While the free spin does not contribute to energy, it modifies the degeneracy with a multiplicative factor. 

As seen in Table.~\ref{tab.degen}, both approaches give the same degeneracy for the ground state. However, for excited states, the two approaches agree to $\mathcal{O}(S)$. As the semi-classical limit is strictly justified for large $S$ spins, we conclude that the degeneracies match. Nevertheless, the $\mathcal{O}(S^0)$ discrepancy is significant. 
It shows that the quadrumer problem ($N=4$) is markedly different from the dimer and the trimer. For both $N=2$ and $N=3$, the appropriate semiclassical description accurately captures the degeneracies in the spectrum. 
For $N=2$, conventional quantum analysis gives eigenvalues $Jj(j+1)\hbar^2$ with $j = 0, 1, \cdots,2S$, with level degeneracy $(2j+1)$. From the semiclassical point of view, this problem maps to a particle on a sphere with the same form of the eigenenergies and degeneracies, except that $j$ runs from $0,\cdots,\infty$.
For $N=3$, the conventional quantum approach gives eigenvalues 
$Jj(j+1)\hbar^2$  where $j = 0, 1,\cdots,3S$. The low lying states, with $j\leq S$, have degeneracy $(2j+1)^2$.
Semiclassically, this problem maps to a spherical top rigid rotor. Once again, this gives the same expressions for the eigenenergies and degeneracies. However, $j$ runs from $0,\cdots,\infty$. In both cases, the low energy spectrum ($j\lesssim S$) is accurately captured by the semiclassical mapping. However, for $N=4$, we find subleading discrepancies in the degeneracy. This could be due to two reasons: \\

(a) \textbf{Order by disorder:} For $N=2,3$, all classical ground states are symmetry-related. As a consequence, quantum fluctuations, arising from terms with lower powers of $S$ in the action, cannot lift their degeneracy.  
However, for $N=4$, ground states with differing values of $(\theta,\phi)$ are not related by any symmetry. This allows quantum fluctuations to have a stronger role. In principle, higher order (lower power in $S$) corrections can induce a preference for certain values of $(\theta,\phi)$ via the well known phenomenon of `order by disorder'. Such corrections could alter the form of the action, for instance, by coupling the rotor and spin degrees of freedom. This could give rise to the observed corrections in the level degeneracies.

A rigorous derivation of $1/S$ corrections is beyond the scope of this study. Nevertheless, we make the following observations.
With regard to the ground state degeneracy, we find perfect agreement between the semiclassical and full quantum results despite the possibility of order by disorder effects. In the semiclassical picture, the ground state degeneracy arises from the emergent free spin while the rotor is in its non-degenerate ground state. This suggests that quantum fluctuations do not play a role when the rotor is in its ground state, leaving the free moment intact. When the rotor is excited, quantum fluctuations could couple it to the free spin, leading to the observed 
corrections in the degeneracies. 

(b) \textbf{Imperfect semi-classical mapping:}
For $N=2$ and $N=3$, the semiclassical large-$S$ path integral precisely reduces to a particle on a sphere and a rigid rotor respectively. However, for $N=4$, the mapping to a rotor and a free spin is approximate due to a subtle effect that arises from the non-trivial topology of the ground state space. This is discussed in detail in the next section. As the mapping itself is only approximate, we can have discrepancies between the semi-classical and quantum spectra. This could also lead to the observed discrepancies in degeneracies. 
\onecolumngrid

\begin{table}
\begin{centering}
\begin{tabular} { | c | c | c | c |} 
 \hline
State & Energy & Degeneracy: Full quantum & Degeneracy: Semi-classical \\ 
 \hline
 \hline
Ground state & 0 & $(2S +1)$ &   $(2S +1)$ \\
First excited & $2J\hbar^2$ & $3^2(2S +1) - 9$ & $3^2(2S +1)$ \\ 
Second excited &$6J\hbar^2$&  $5^2(2S +1)-45$& $5^2(2S +1)$ \\
\vdots   & \vdots       &\hspace{0.4cm} \vdots  &\hspace{0.4cm} \vdots  \\
 \hline
\end{tabular}
\end{centering}
\caption{Quadrumer spectrum from `full quantum' and semi-classical approaches (see text). Both approaches give the same energies, shown in the second column. The degeneracy from the two approaches is shown in the third and fourth columns. }
\label{tab.degen}
\end{table}

\twocolumngrid

\section{Singularities in the ground state space}
There is a non-trivial subtlety in the identification of $\mathcal{Z}_2$ in Eq.~\ref{eq.partitionfunction} as the path integral of a rigid rotor. The rotor angular momentum is given by $\vec{L}' = R M\vec{L}$. The $3\times 3$ matrix $M$ here depends on the variables $\theta$ and $\phi$. For generic values of $\theta$ and $\phi$, $M$ has three non-zero eigenvalues. This leads to three independent components of $\vec{L}'$, as required to describe the angular momentum of a rigid rotor. However, the matrix $M$ becomes singular at three isolated values of $(\theta,\phi)$ at which the spin configuration is collinear. At these points, one of the eigenvalues of $M$ vanishes, leaving us with only two degrees of freedom in $\vec{L}'$; it can no longer be identified as the angular momentum of a rigid rotor. 
Strictly speaking, this forbids the identification of $Z_2$ in Eq.~\ref{eq.partitionfunction} with a rigid rotor.

This effect originates from the parametrization in Eq.~\ref{eq.S_parametrization}. Suppose all four unit vectors, $\hat{n}_j$'s, are collinear and aligned along $\pm\hat{z}$. In this case, the $z$-component of $\vec{L}$ becomes redundant in Eq.~\ref{eq.S_parametrization}. Note that it is the projection of $\vec{L}$ onto the plane perpendicular to $\hat{n}_j$ that enters $\vec{S}_j$. With all spins parallel to $\hat{z}$, we can assign any value to $L_z$ without changing any of the spins. This can be understood by visualizing all possible small fluctuations about a collinear state. The system can only develop a non-zero magnetization in two directions, while preserving the length of each spin. These form the two independent components of $\vec{L}'$.

While our semiclassical mapping fails at collinear states,  this is nevertheless a minor effect as the number of such ground states is very small. In fact, collinear ground states form a set of measure zero as they occur for three isolated values of $(\theta,\phi)$. By neglecting this set in the integration space of Eq.~\ref{eq.partitionfunction}, we can 
persist with our identification of the system with a spin-$S$ spin and a rigid rotor. 
That this is a good approximation can be seen from the excellent agreement in the spectrum as shown in Tab.~
\ref{tab.degen}.

\section{Soft fluctuations}

To understand the topology of the ground state space, it is instructive to look at `soft' fluctuations. Given a ground state configuration of the four spins in the cluster $\{\vec{S}_j\}$, we consider small deviations of the spins, $\{\delta \vec{S}_j \}$ with $j=1,\ldots,4$. As we have eight degrees of freedom in total (2 per spin), we always have eight independent fluctuation modes labeled as $\{\delta \vec{S}_j^\alpha \}$, with $\alpha=1,\ldots,8$. We identify independent modes using the condition $\sum_{j=1}^N \{\delta \vec{S}_j^{\alpha} \cdot \delta \vec{S}_j^{\beta} \} = \delta_{\alpha,\beta}$.

These small fluctuation modes can be naturally classified as `hard' and `soft'. Hard modes take us out of the ground state space -- they induce a net magnetization in the cluster, i.e., $\sum_j \{\delta \vec{S}_j^{\alpha}\} \neq 0$. In contrast, soft modes preserve the zero-total-spin condition and keep us within the ground state space. 

\begin{figure*}
\includegraphics[height=41mm, width=37mm]{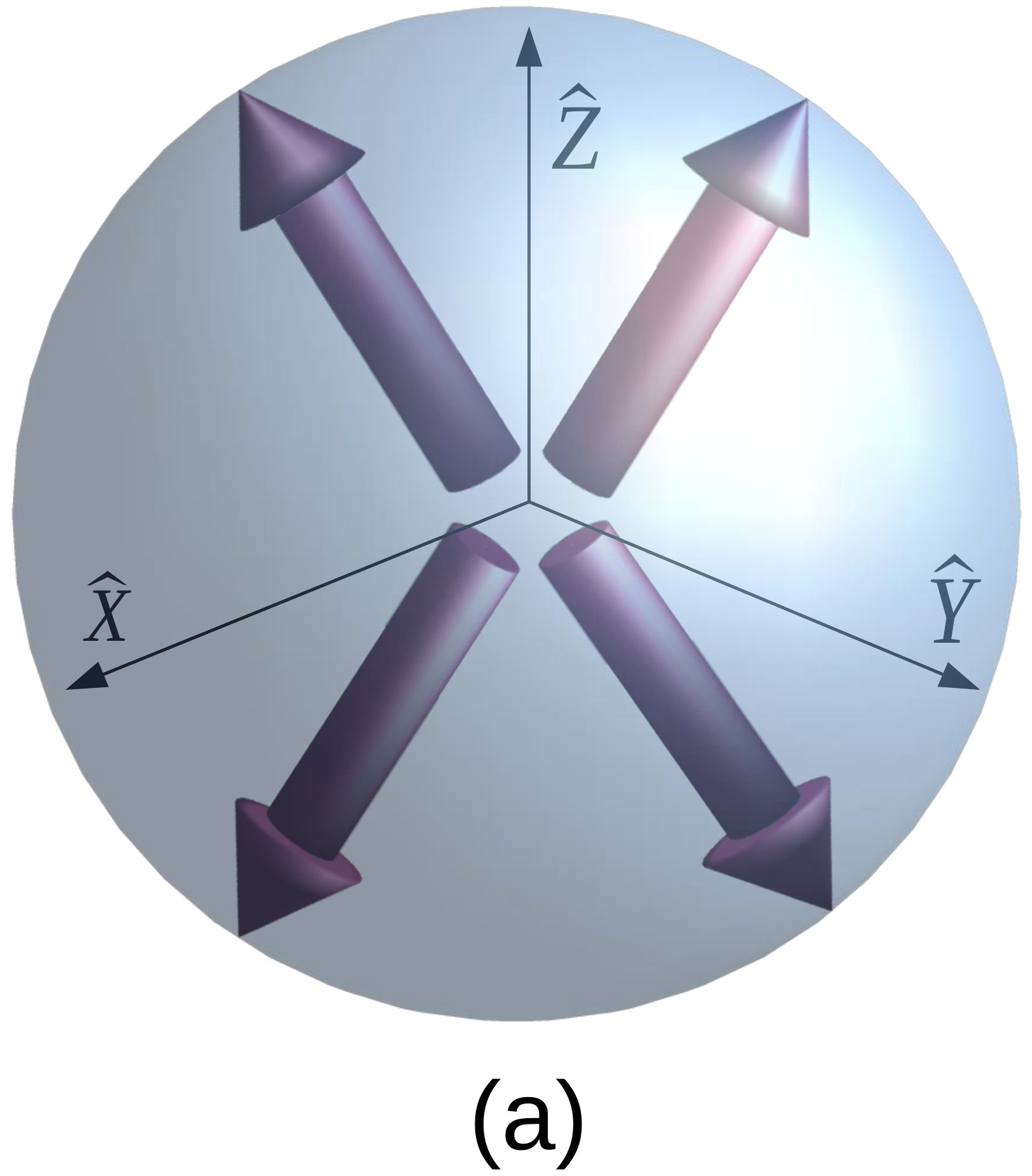}\hspace*{0.25 in}
\includegraphics[height=41mm, width=37mm]{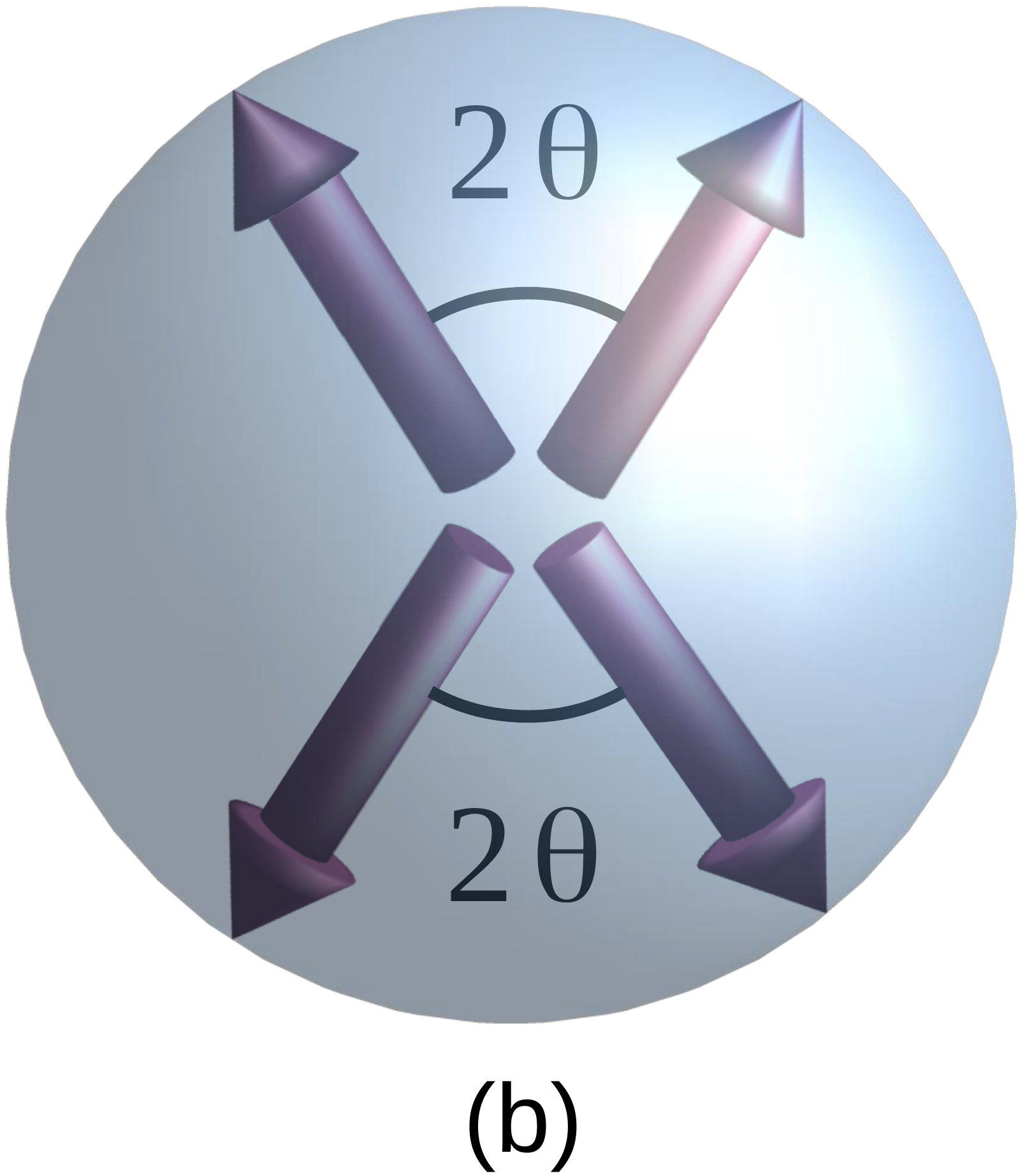}\hspace*{0.1 in}
\includegraphics[height=41mm, width=37mm]{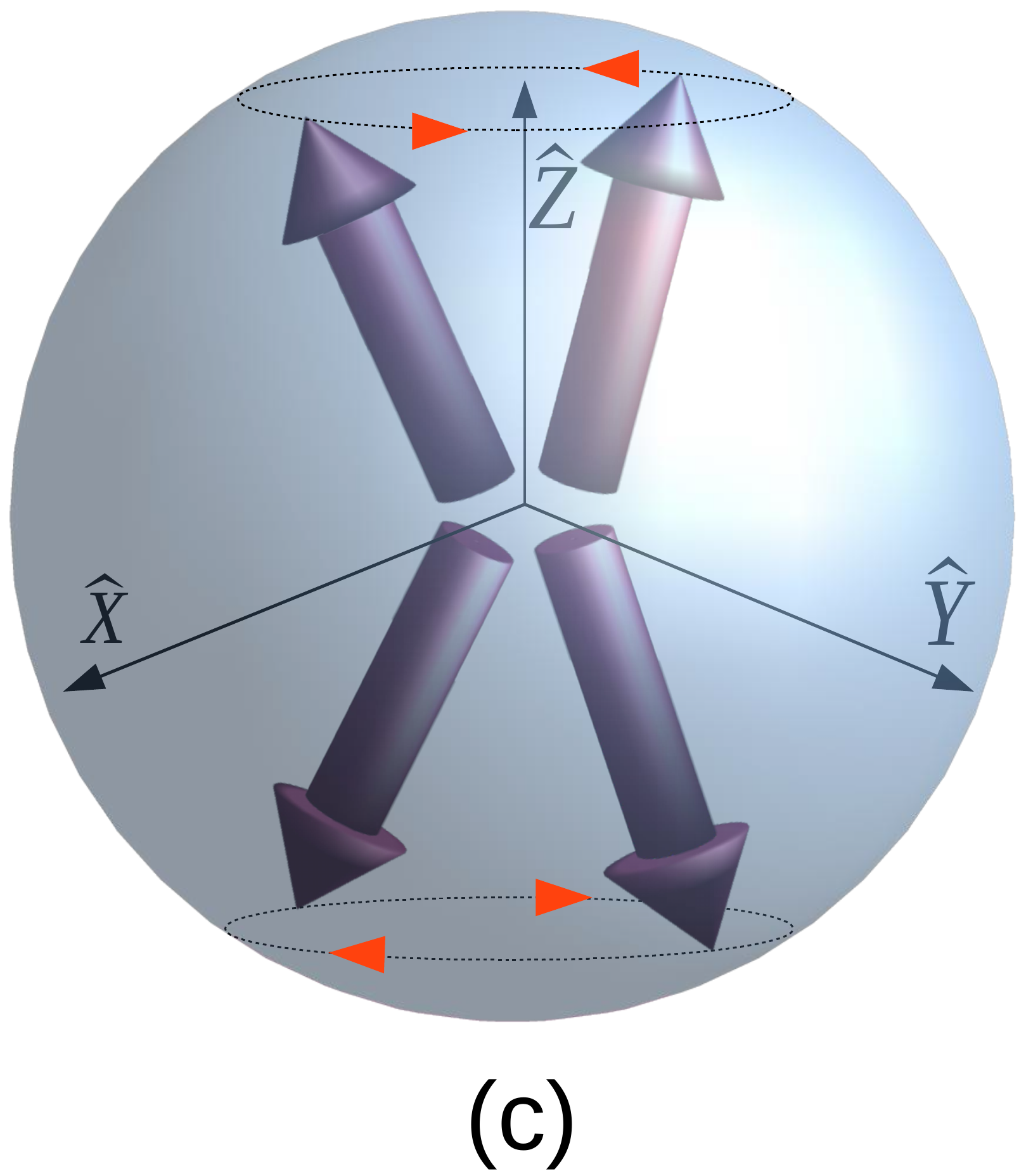}\hspace*{0.1 in}\\
\vspace{0.1 in}
\includegraphics[height=41mm, width=37mm]{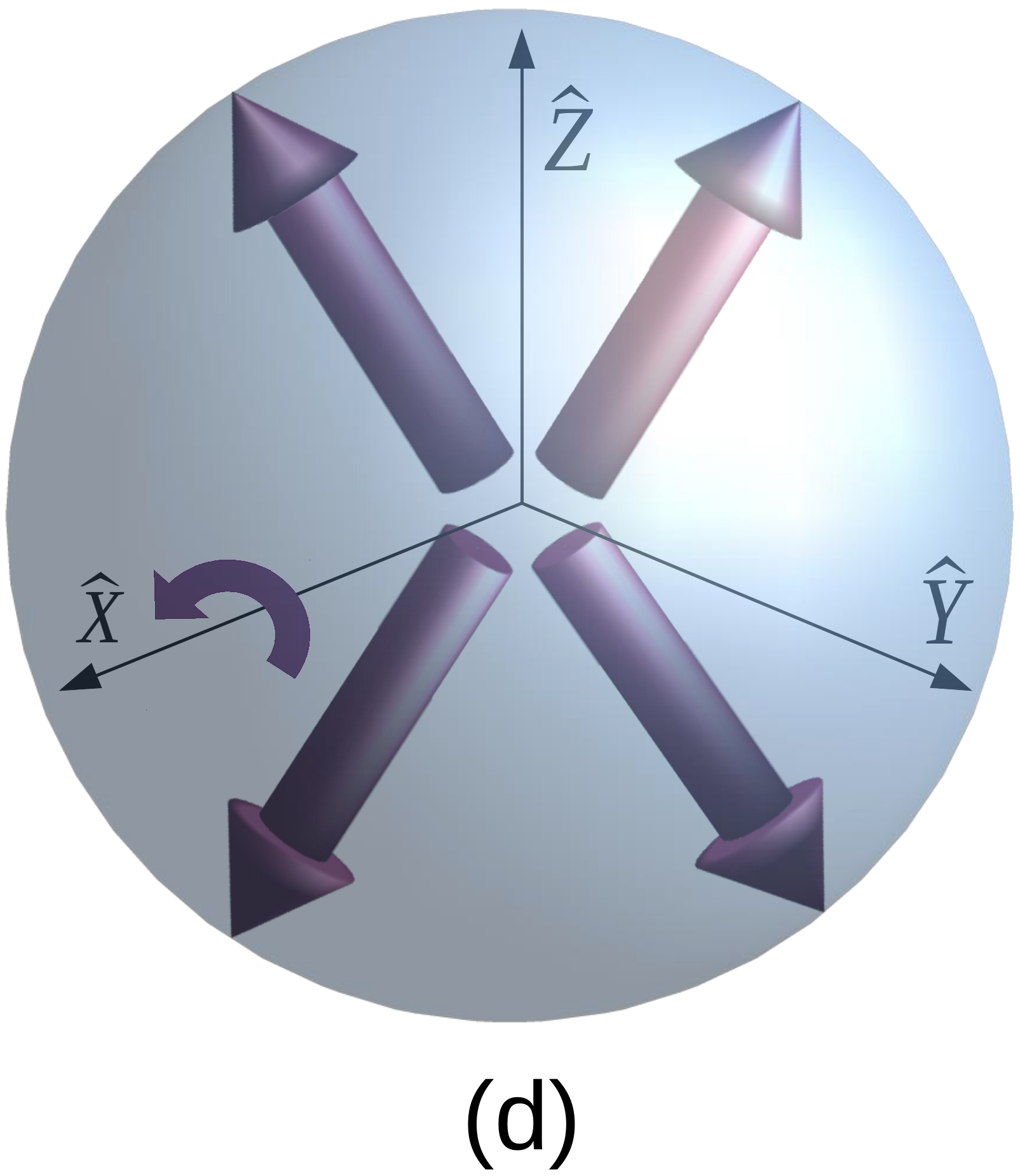}\hspace*{0.1 in}
\includegraphics[height=41mm, width=36mm]{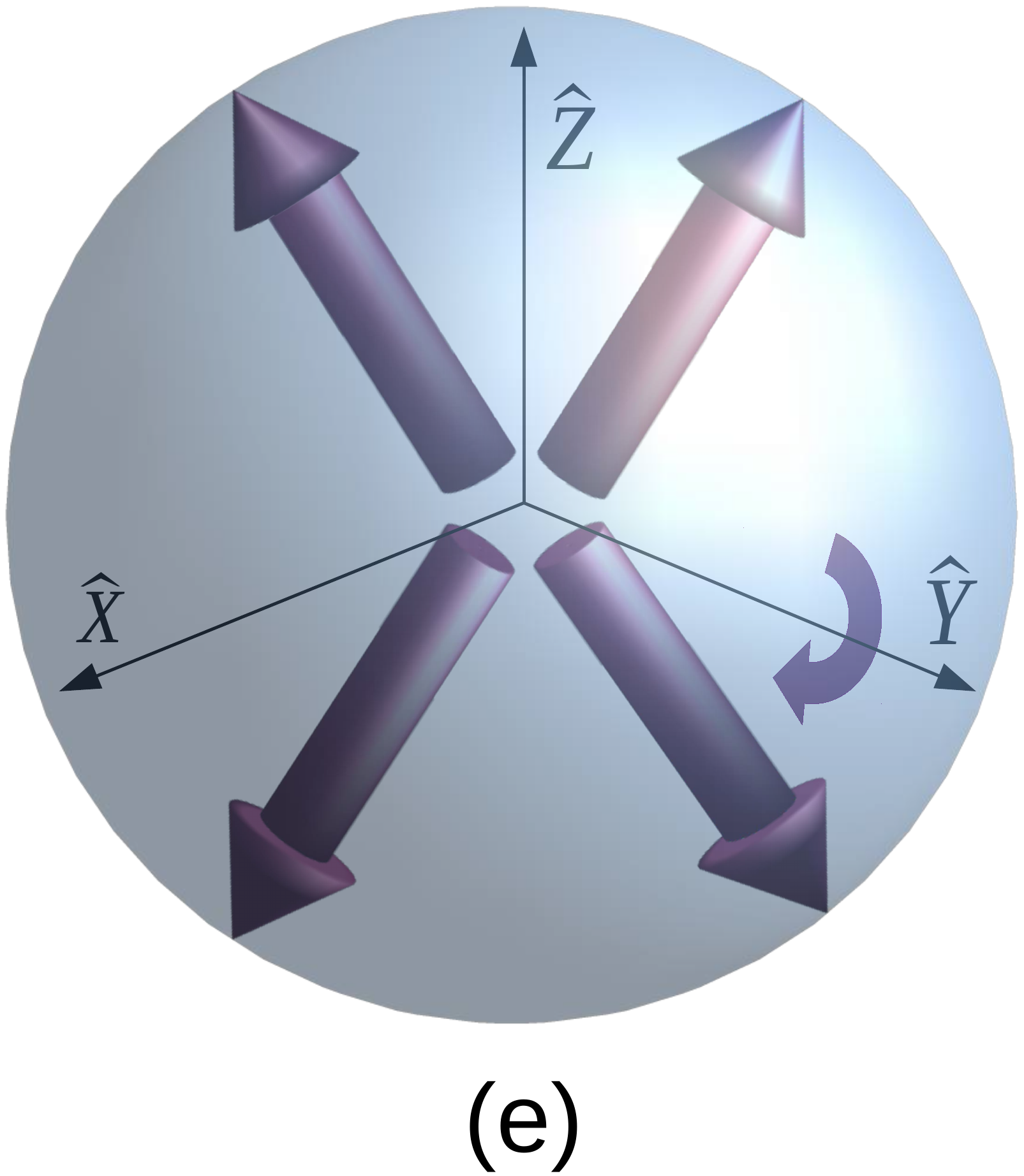}\hspace*{0.1 in}
\includegraphics[height=41mm, width=36mm]{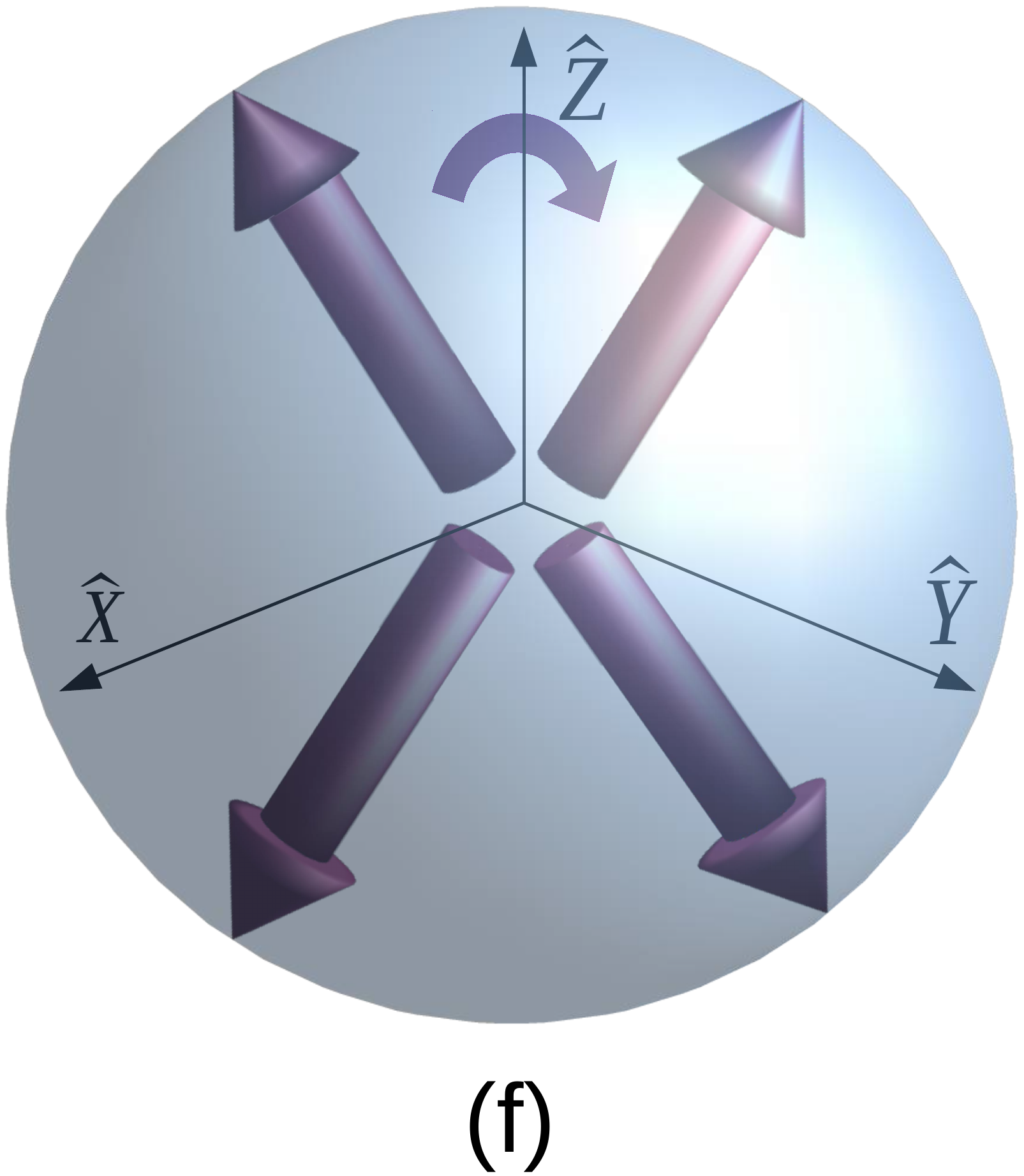}\hspace*{0.1 in}

\caption{Soft fluctuations about a coplanar state: (a) A reference coplanar state, (b) a variation in $\theta$, changing the angular separation between spins, (c) variation in $\phi$, changing the twist angle between planes of pairs of spins, (d, e, f) rigid rotations about the $X$, $Y$, and $Z$ axes. These are five independent soft modes. }
\label{fig2}
\end{figure*} 

\begin{figure*}
\includegraphics[height=41mm, width=37mm]{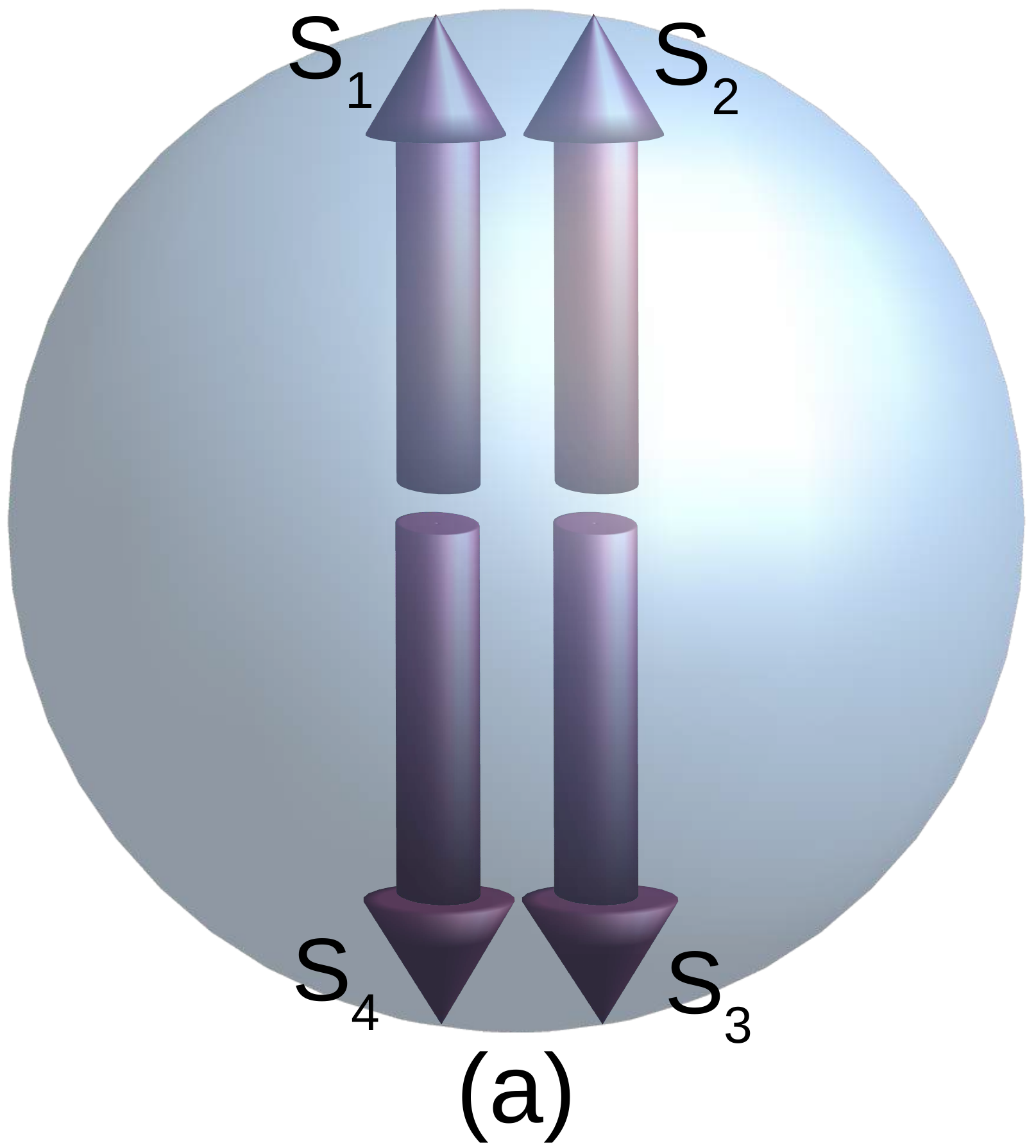}\hspace*{0.25 in}\includegraphics[height=41mm, width=37mm]{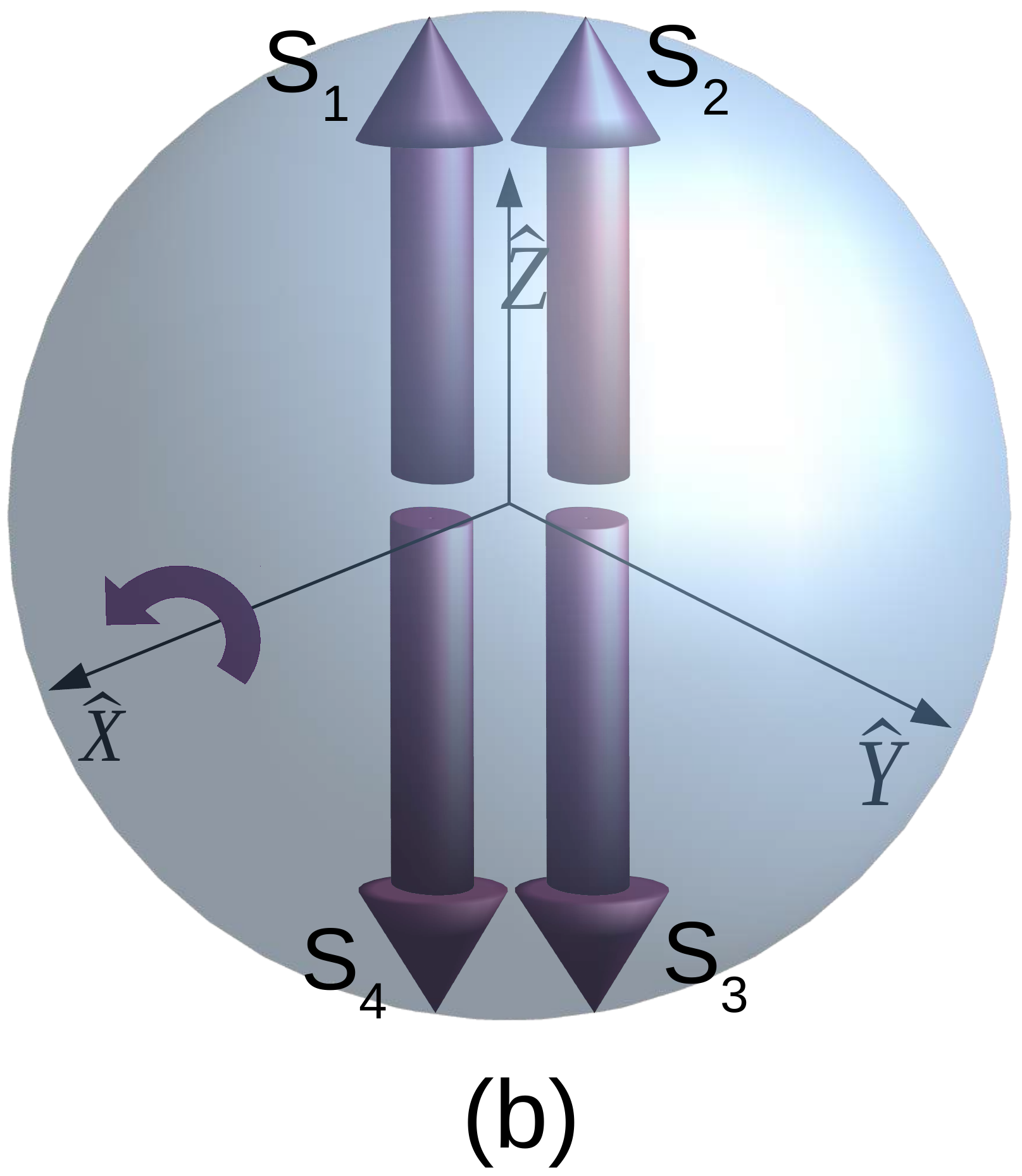}\hspace*{0.1 in}\includegraphics[height=41mm, width=36mm]{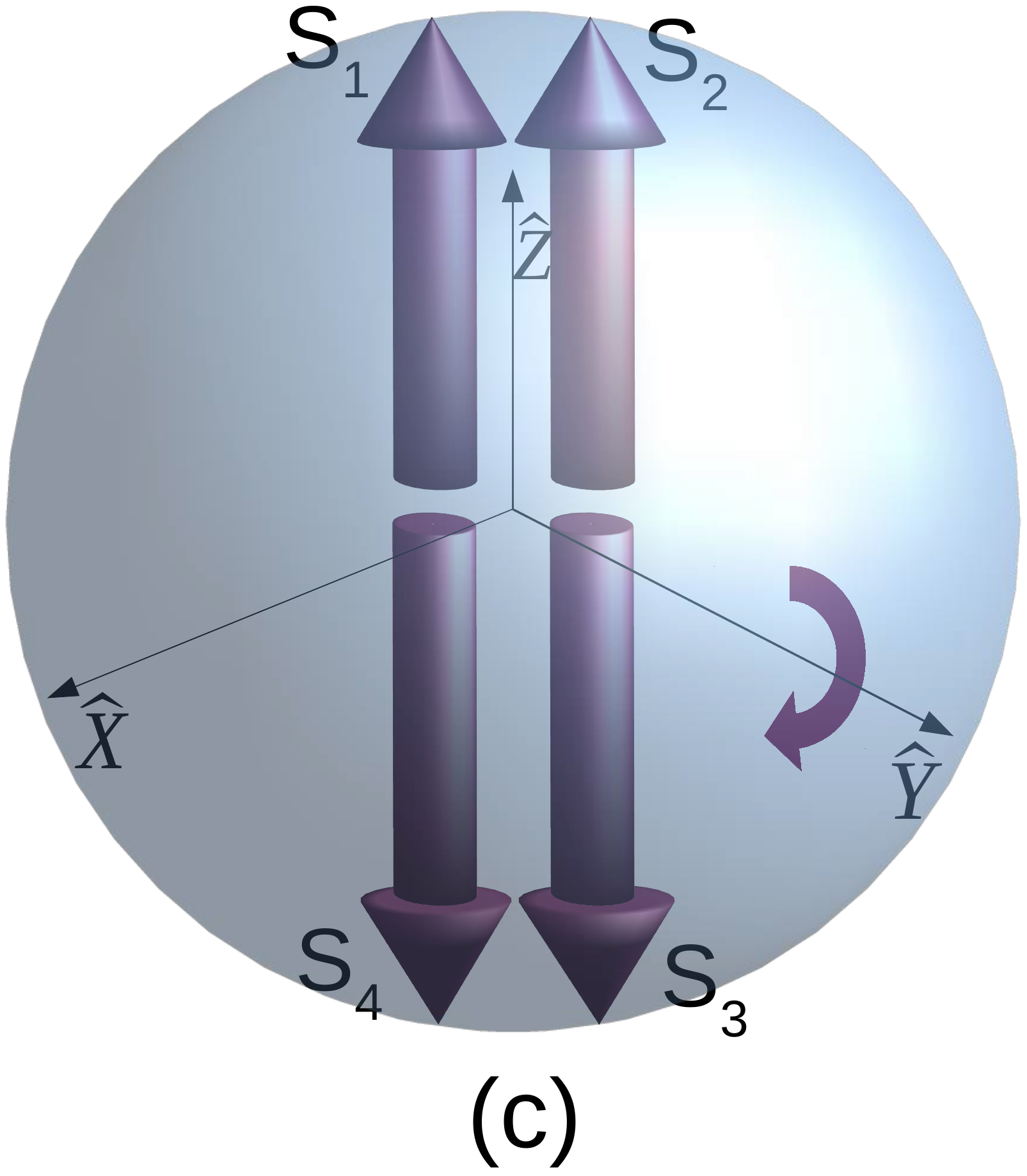}\hspace*{0.1 in}
\\
\vspace{0.1 in}
\includegraphics[height=41mm, width=36mm]{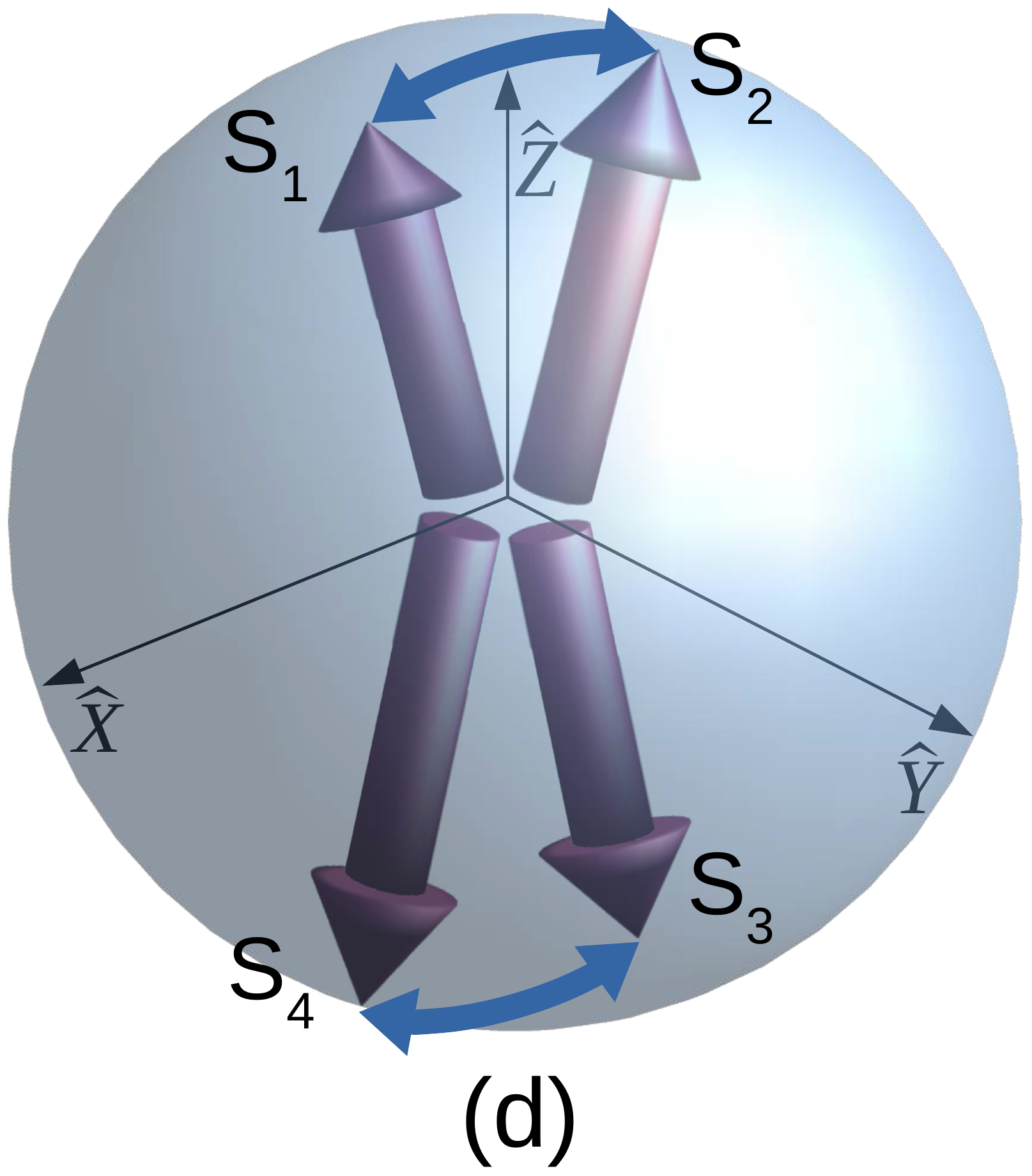}\hspace*{0.1 in}
\includegraphics[height=41mm, width=37mm]{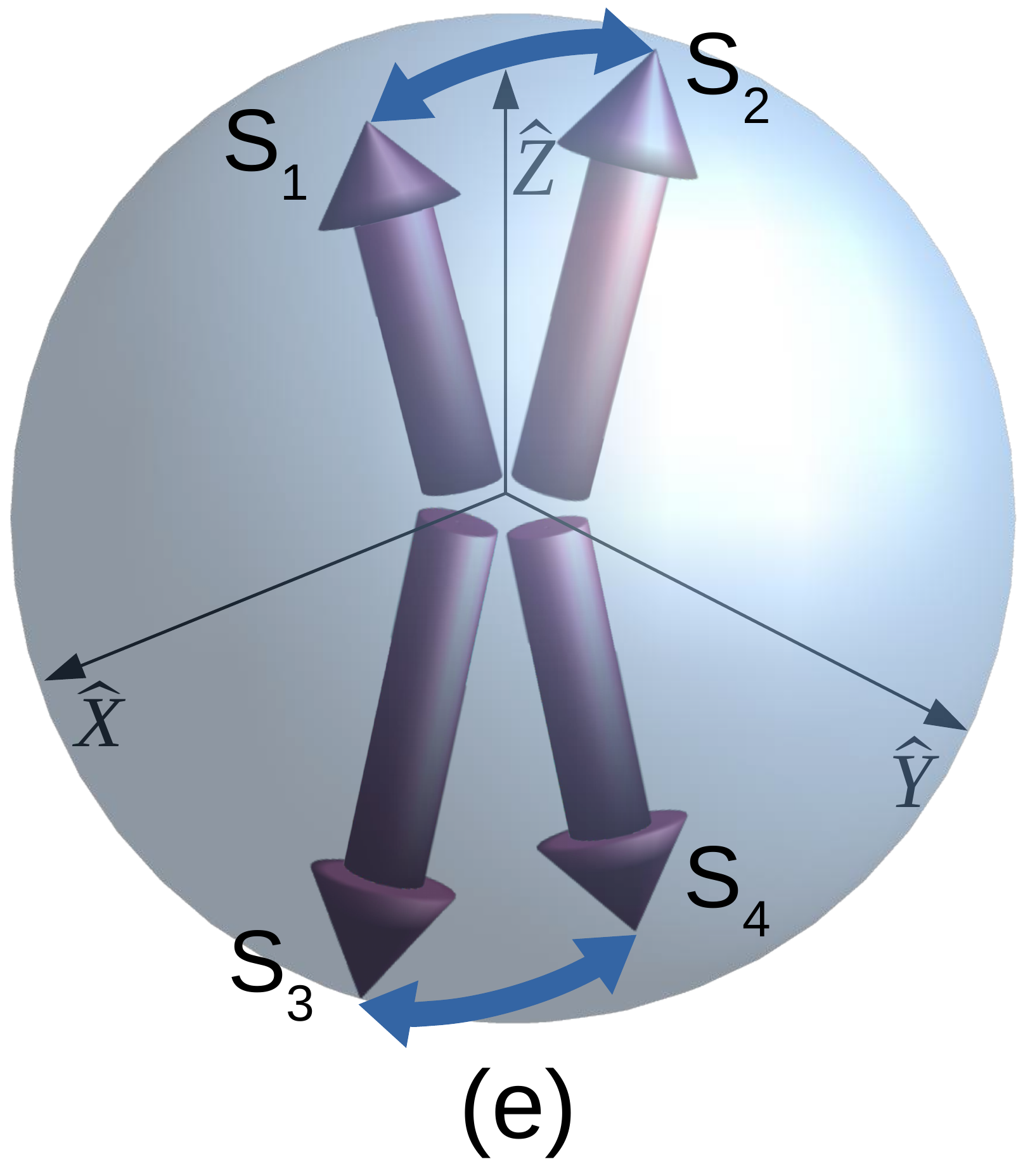}\hspace*{0.1 in}
\includegraphics[height=41mm, width=36mm]{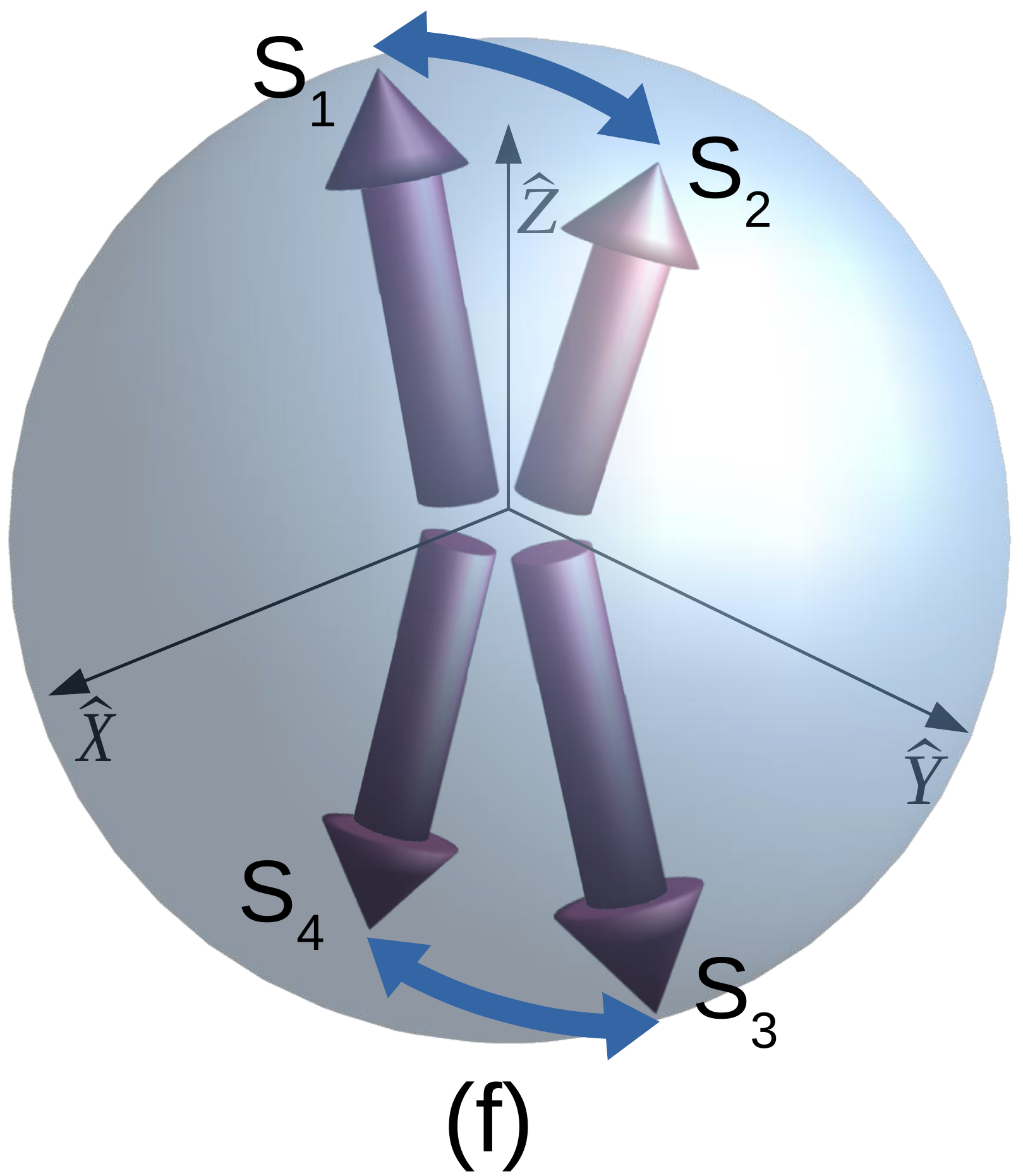}\hspace*{0.1 in}
\includegraphics[height=41mm, width=37mm]{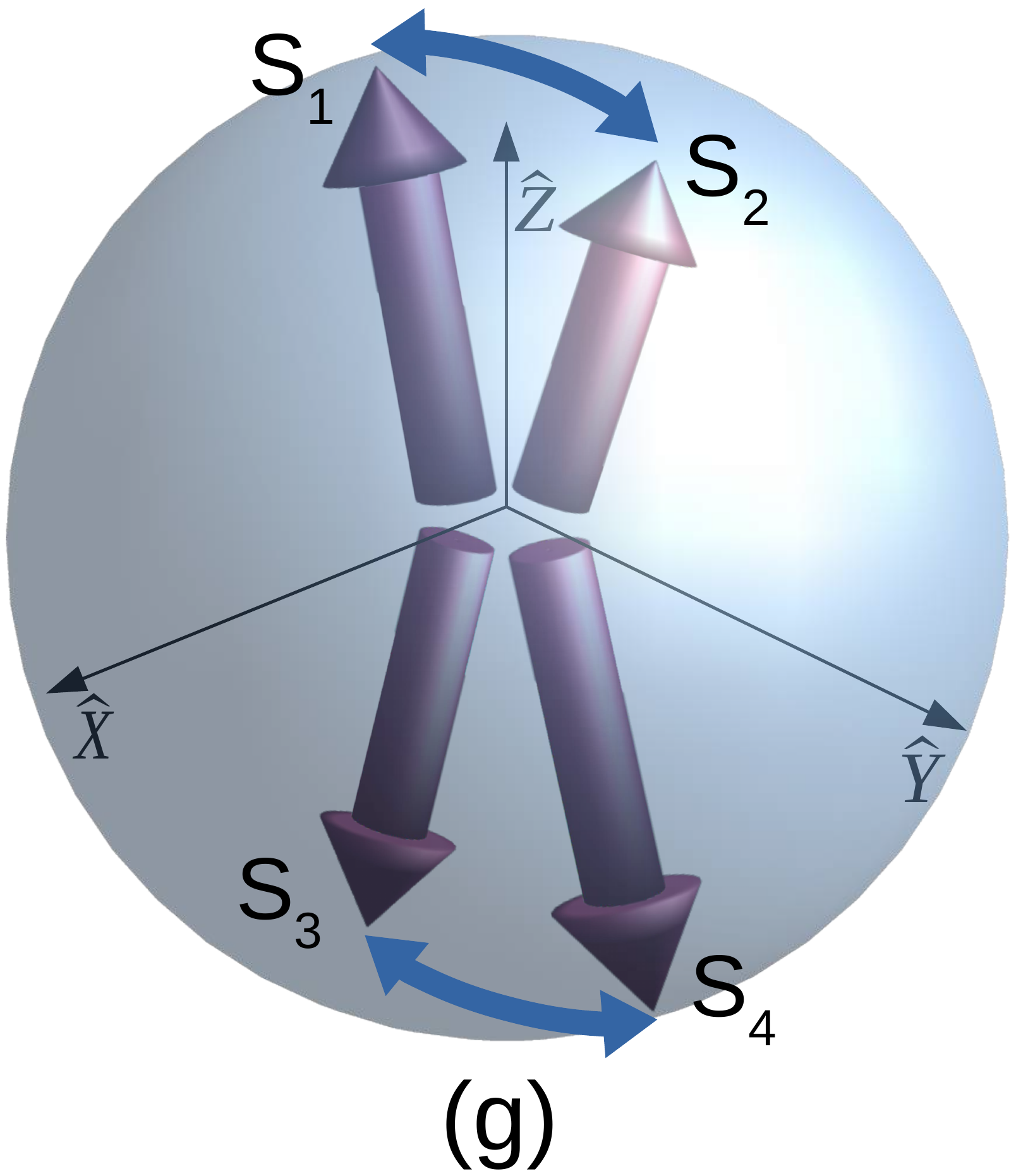}\hspace*{0.1 in}
\caption{Soft fluctuations about a collinear state: (a) A reference collinear state with moments aligned along the $Z$ axis, (b, c) rigid rotations about the $X$ and $Y$ axes, (d) a distortion in the $XZ$ plane, (e) a similar distortion but with $\vec{S}_3$ and $\vec{S}_4$ switched, (f) a distortion in the $YZ$ plane, and (g) a similar distortion but with $\vec{S}_3$ and $\vec{S}_4$ switched. These form six independent soft modes.}
\label{fig3}
\end{figure*} 

In Fig.~\ref{fig2} and Fig.~\ref{fig3}, we pictorially depict the soft modes around (a) a coplanar state and (b) a collinear state respectively. There are five soft modes about the coplanar state corresponding to varying $\theta$, varying $\phi$ and three independent rotations. Indeed, this is true of \textit{all non-collinear} ground states; each state allows for five soft modes which can be understood in the same way. However, collinear states allow for \textit{six} soft modes as shown in Fig.~\ref{fig3}; see App.~\ref{app.collinearsoft} for explicit expressions. These correspond to two independent rotations and four independent deformations.

Mathematically, the set of soft modes describes the tangent space around a given element of the ground state space. For the ground state space to be a manifold, the tangent space at every point must have the same dimensionality. To be precise, the neighbourhood of every point must be isomorphic to $\mathbb{R}^n$, where $n$ is the dimension of the space. 
Here, we have an extra sixth dimension whenever the state is collinear.
We assert that this demonstrates a deep mathematical property, viz., the non-manifold character of the ground state space. 
In Appendix~\ref{implicit.func.proof}, we provide a rigorous proof that the ground state space, \textit{with collinear configurations removed}, forms a five dimensional manifold. We show this using the implicit function theorem which provides a sufficient condition for manifold character.

To illustrate the singularities that occur at collinear states, we discuss a `spin-wave approach' in Appendix~\ref{spinwave}. Assuming that the cluster always remains in the vicinity of a collinear state, we develop a path integral description. With only two rotational degrees of freedom, the system maps to a rigid \textit{rod} rather than a rigid body.

\section{Thermodynamics}

We have shown that the semi-classical description succeeds in describing the low-energy spectrum of the quadrumer. We can now reinterpret thermodynamic properties as arising from the semi-classically obtained free spin-$S$ moment and rigid rotor.  
From this point of view, the partition function of the quadrumer is given by 
\begin{eqnarray}
\mathcal{Z} = \sum_{j}(2S+1)(2j+1)^2e^{-j(j+1)\beta J\hbar^2}.
\label{eq.Zexp}
\end{eqnarray} 
At zero temperature, the quadrumer has non-zero entropy, $k_B\ln(2S+1)$, arising from the degeneracy of the free spin-$S$ spin. This non-vanishing entropy can be seen in specific heat measurements on candidate materials. 

To find low temperature properties, we may retain the first few $j$ values in Eq.~\ref{eq.Zexp}. For example, retaining only $j=0,1$, the entropy can be approximated as $S_{ent} \approx k_{B}\left[\ln(2S+1) + \ln(1 + 9e^{-2\beta J\hbar^2}) + \frac{18 J\beta\hbar^2}{e^{2J\beta\hbar^2} + 9} \right]$, where the first term is the free spin contribution and the rest is the rigid rotor contribution. Similarly, the specific heat at low temperatures comes out to be $C_v = 36 k_B (J\beta\hbar^2) \frac{e^{2\beta J\hbar^2}}{(9 + e^{2\beta J\hbar^2})^2}$. The free spin does not contribute to specific heat as it does not contribute to energy.

Our formalism also allows us to directly calculate magnetic susceptibility. Assuming a small magnetic field of strength $B$ along $\hat{z}$, it enters the Hamiltonian as a new term, $-B {L'}_z$, where $L'_z$ is the $z$ component of the magnetization. The partition function changes to $\mathcal{Z} = \sum_{j}(2S+1)(2j+1)e^{-j(j+1)\beta J\hbar^2}\left(\sum_{m=-j}^j e^{m\beta B\hbar}\right)$, where the eigenvalues of $L'_z$ are given by $m\hbar$. The low temperature susceptibility comes out to be $\chi = \frac{1}{\beta}\partial_B^2 {\ln \mathcal{Z}}|_{B\rightarrow 0}= \frac{6\beta\hbar^2}{e^{2\beta J\hbar^2} + 9}$, in agreement with known results in the large-$S$ limit\cite{Garcia2000,Garcia2001}. The free spin does not contribute to the susceptibility as well. As the emergent spin only signifies internal coordinates (relative angles between spins), it does not couple to an external magnetic field.

\section{Summary and discussion}
We have presented a path integral description for the quantum spin quadrumer. The ground state space of this system has non-trivial topology, reflected in the differing number of soft fluctuations around collinear and non-collinear ground states. This provides a simple and an experimentally realizable example of dynamics on a non-manifold space. Earlier studies of the quadrumer with Dzyaloshinskii-Moriya couplings have discussed one-dimensional ground state spaces with non-manifold character\cite{Elhajal2005,Canals2008}. Our study with purely Heisenberg couplings brings out a larger five-dimensional non-manifold space.

We have shown  that the quadrumer decouples into a free spin-$S$ spin and a rigid rotor. This provides a beautiful example of `emergence' -- the internal spin configuration manifests as a spin-$S$ order parameter in the low energy description. This spin character does not appear directly in the microscopic description; it cannot be deduced from a conventional study of the cluster and its dynamics. Indeed, although the quadrumer has been extensively studied, this property has not been brought out so far. An early indication of an emergent spin may be found in the equations for semiclassical dynamics derived in Ref.[\cite{PhysRevLett.81.5221}].

We have considered a spin-$S$ cluster in which every spin is coupled to every other spin. This structure arises naturally in tetrahedral molecules/clusters which have four spins at the corners of a regular tetrahedron. Several experimental realizations are known to exist. The canonical examples are transition metal tetrahedra\cite{Jian2006}. Notably, Ni$_4$Mo$_{12}$ realizes a near-perfect tetrahedron of $S=1$ moments\cite{Mller2000,Nehrkorn2010}. A $S=1/2$ tetramer is also realized in ${\text{Cu}}_{4}{\text{OCl}}_{6}{\text{daca}}_{4}$\cite{Zaharko2008}. More recently, several lanthanide-based compounds have been synthesized. A Dy-based molecular magnet\cite{Lin2011} realizes a $S=5/2$ tetrahedron, but with Ising-like anisotropy. These molecular magnets closely resemble our problem. Most of their experimental properties can be explained with a conventional quantum analysis of four spins (e.g., as in Ref.~\onlinecite{Schnack2000}). Our work reinterprets these as emanating from an emergent rigid rotor and a free spin.

An exciting development is the synthesis of `breathing' pyrochlore magnets with weakly coupled tetrahedra. An example from the lanthanide family is Ba$_3$Yb$_2$Zn$_5$O$_{11}$ in which each Yb atom forms a pseudospin-$1/2$ moment\cite{Kimura2014,Haku2016,Rau2016,Park2016}. The inter-tetrahedron coupling is so weak that the tetrahedra are essentially isolated; this is reflected in the neutron scattering intensity showing flat momentum-independent modes. 
Our analysis may not be directly applicable here, due to the presence of strong Dzyaloshinskii-Moriya interactions. Likewise, LiGaCr$_4$O$_8$ and LiInCr$_4$O$_8$ form breathing pyrochlores with Cr$^{3+}$ $S=3/2$ moments. They vary in the degree of breathing and show intriguing ordering properties\cite{Hiroi2014}. Theoretical proposals have been put forward to explain their ordering\cite{Koga2001,Tsunetsugu2001,Tsunetsugu2017}. Our results will help to develop a field theoretic description for these systems. The 
additional couplings in these materials such as anisotropies, dipolar interactions, etc., will modify our semi-classical picture and couple the rotor and the spin fields.

The real promise of our study is that it provides a starting point for semi-classical field theories. The $N=4$ cluster is the building block of pyrochlore magnets, the checkerboard lattice\cite{Canals2002,Brenig2002,Bishop2012,Moura2017, Fouet2003,Hermele2004,Bernier2004}, the four-leg tube\cite{Plat2015}, and the square J$_1$-J$_2$-J$_3$ antiferromagnet\cite{Danu2016}. 
In particular, the Heisenberg pyrochlore magnet is of great interest as a canonical model of frustration with realizations in spinel compounds\cite{Reimers1991,Moessner1998,Tsunetsugu2002,Yamashita2000,DelMaestro2004,Willis2006,
Henley2005, Henley2006, Hizi2007, Conlon2009, Lapa2012, Berg2003}. 
Its ground state and ordering properties continue to be debated\cite{Canals1998, Tsunetsugu2001PRB,Huang2016}.
With new experimental realizations emerging\cite{Higo2017}, we hope that a suitable field theoretic approach will throw light on this model and its intriguing properties.

\acknowledgments We thank Pralay Chatterjee and Arghya Mondal for many helpful discussions about the mathematics of manifolds. We thank Karlo Penc, Srivatsa N. S., B. Sathiapalan and Yuan Wan for useful discussions and comments.

\appendix
\section{Berry Phase}\label{appA}
For a single spin, the first term in the Berry phase in Eq.~\ref{eq7} can be written in the following way \cite{Berry_phase},
\begin{align}
&S_B = -S\int_0^\beta d\tau\; tr\left( \partial_\tau U U^\dagger \sigma^z \right), 
\end{align}
where $U$ is a matrix that encodes the $j^{th}$ spin. The coordinates of the spin, in the reference frame, are denoted by polar angle $\tilde{\theta}_j$ and azimuthal angle $\tilde{\phi}_j$. This moment is rotated by the rotation matrix $R$.
The matrix $U$ is defined so as to transform $\hat{z}$ to $\hat{n}_j=\hat{n}(\tilde{\theta}_j, \tilde{\phi}_j)$, followed by a global rotation to give $R\hat{n}$.
This operation is given below.
\begin{align}
&U_R^\dagger (e^{-i\frac{\tilde{\phi}_j}{2}\sigma^z})(e^{-i\frac{\tilde{\theta}_j}{2}\sigma^y})\sigma^z (e^{i\frac{\tilde{\theta}_j}{2}\sigma^y})(e^{i\frac{\tilde{\phi}_j}{2}\sigma^z})U_R.
\end{align}
Here, $U = e^{i\frac{\tilde{\theta}_j}{2}\sigma^y}e^{i\frac{\tilde{\phi}_j}{2}\sigma^z}U_R = U(\tilde{\theta}_j, \tilde{\phi}_j) U_R$, where $U_R$ is the unitary transformation corresponding to rotation and $\sigma$'s are Pauli matrices.
\begin{align}\label{eq16}
\mbox{Now,}&\;\; \tr\left( \partial_\tau U U^\dagger \sigma^z \right) \nonumber\\
&= \tr\left( \left\{ \partial_\tau \left(U(\tilde{\theta}_j,\tilde{\phi}_j)U_R\right)\right\} \left(U_R^\dagger U^\dagger(\tilde{\theta}_j,\tilde{\phi}_j) \right)\sigma^z\right)\nonumber\\
& =\tr\left(\partial_\tau U(\tilde{\theta}_j,\tilde{\phi}_j) U^\dagger(\tilde{\theta}_j,\tilde{\phi}_j) \sigma^z\right) \nonumber\\
&\hspace{0.7cm}+ \tr\left(\partial_\tau U_R U_R^\dagger U^\dagger(\tilde{\theta}_j,\tilde{\phi}_j) \sigma^z U(\tilde{\theta}_j,\tilde{\phi}_j)\right).
\end{align}
It can be easily checked that $U^\dagger(\tilde{\theta}_j,\tilde{\phi}_j) \sigma^z U(\tilde{\theta}_j,\tilde{\phi}_j) = \hat{n}_j\cdot\sigma$. The second term in Eq.~\ref{eq16} becomes \hspace{0.1cm}$\tr\left(\partial_\tau U_R U_R^\dagger \hat{n}_j\cdot\sigma\right)$.
So far, we calculated the Berry phase term for a single spin. Adding the contributions from 4 spins, we get
\begin{align}
&\sum_{j=1}^4\tr(\partial_\tau U_R U_R^\dagger \hat{n}_j\cdot\sigma)\nonumber\\
 &= \tr(\partial_\tau U_R U_R^\dagger \sum_j\hat{n}_j\cdot\sigma)\nonumber\\
&= 0.
\end{align}
This follows from the ground state condition, $\sum_j\hat{n}_j=0$.
The first term in Eq.~\ref{eq16} can be shown to be $(i\,\cos\tilde{\theta}_j \, \dot{\tilde{\phi}}_j )$. 
\begin{equation}
S_B  = -iS\int_0^\beta d\tau \cos\tilde{\theta}_j\,\dot{\tilde{\phi}}_j,
\end{equation}
this term is again for a single spin with $\tilde{\theta}_j$ and $\tilde{\phi}_j$ being the polar and azimuthal angles. We now add contributions from four spins with the corresponding polar and azimuthal angles: $(\theta,\frac{\phi}{4}), (\theta,\pi +\frac{\phi}{4}), (\pi - \theta,\pi-\frac{\phi}{4})$ and $(\pi - \theta, 2\pi-\frac{\phi}{4})$:
\begin{align}
S_B^{\mathrm{four\phantom{a} spins}} = \sum_{j = 1}^4 S_B^j= -iS\int_0^\beta d\tau  \cos\theta\,\dot{\phi},
\end{align}
where $\theta$ and $\phi$ are the parameters used to describe a generic ground state in Fig.~\ref{fig1}.

\begin{widetext}
\section{Jacobian derivation}\label{appB}
At any given imaginary time slice, the path integral measure takes the form 
\begin{align}\label{eqB1}
\int\, d\Omega_{1x}\, d\Omega_{1y}\, d\Omega_{1z}\, d\Omega_{2x}\,d\Omega_{2y}\,d\Omega_{2z}\,d\Omega_{3x}\,d\Omega_{3y}\,d\Omega_{3z}d\Omega_{4x}\,d\Omega_{4y}\,d\Omega_{4z} \,\delta (\Omega_1^2 -1)\,\delta(\Omega_2^2 -1)\, \delta(\Omega_3^2 -1)\,\delta(\Omega_4^2 -1).
\end{align}
Let us rewrite the spin parametrization of Eq.~\ref{eq3} in the following way,
\begin{align}
&\hat{\Omega}_1 = R\left(\vec{n} + M_1\frac{\vec{L}}{S}\right)\nonumber\\
&\hat{\Omega}_2 =R\left(\rho T_1\vec{n} + M_2\frac{\vec{L}}{S}\right)\nonumber\\
&\hat{\Omega}_3 = R\left(\sigma T_2\vec{n} + M_3\frac{\vec{L}}{S}\right)\nonumber\\
 &\hat{\Omega}_4 = R\left(\lambda T_3\vec{n} + M_4\frac{\vec{L}}{S}\right), \label{eqB2}
\end{align}
where $\vec{n} = (n_x, n_y, n_z)$. This vector $\vec{n}$ has polar angle $\theta$ and azimuthal angle $\phi/4$, where $\theta$ and $\phi$ are the angles that parametrize the ground state space as shown in Fig.~\ref{fig1}.
As described in the main text, $\hat{\Omega}_{1,\ldots,4}$ are taken to be normalized to $\mathcal{O}(S^{0})$.
We have introduced three new scalar variables, $\rho, \sigma $ and $\lambda$, in order to have twelve new variables. As we now have the same number of old and new variables, we can evaluate the Jacobian of the transformation. We have used $T_1 = diag[-1,-1,1], T_2 = diag[-1,1,-1]$ and $T_3 = diag[1,-1,-1]$. In terms of the new variables, the integral~\ref{eqB1} becomes 
\begin{align}\label{eq22}
\int\, \mathcal{J}_1(\rho,\sigma,\lambda,\gamma,\alpha, \beta,\vec{L},\vec{n})\,d\gamma\,d\alpha\,d\beta\,d\rho\,d\sigma\,d\lambda\,d\vec{L}\,d\vec{n} \,\,\delta(\rho^2 -1)\,\delta(\sigma^2 -1)\,\delta(\lambda^2 -1)\,\delta(n^2 -1).
\end{align} 
Here $\alpha, \beta $ and $\gamma$ parametrize the rotation matrix $R$. $(\alpha,\beta)$ define an axis of rotation, while $\gamma$ represents the rotation angle. After finding the Jacobian $\mathcal{J}_1$, we can integrate out $\rho,\sigma$ and $\lambda$ by replacing $\rho =\sigma =\lambda =1$ in the Jacobian.
\setcounter{MaxMatrixCols}{20}
The Jacobian matrix can be written in the following form,
\begin{align}\label{eqB5}
 \begin{bmatrix}
 \pdv{\hat{\Omega}_1}{\gamma}& \pdv{\hat{\Omega}_1}{\alpha}& \pdv{\hat{\Omega}_1}{\beta}& \pdv{\hat{\Omega}_1}{\rho}& \pdv{\hat{\Omega}_1}{\sigma}& \pdv{\hat{\Omega}_1}{\lambda}&  \pdv{\hat{\Omega}_1}{n_x} & \pdv{\hat{\Omega}_1}{n_y} & \pdv{\hat{\Omega}_1}{n_z} & \pdv{\hat{\Omega}_1}{L_x}& \pdv{\hat{\Omega}_1}{L_y} & \pdv{\hat{\Omega}_1}{L_z}
 \vspace{0.2cm} \\ 
 \pdv{\hat{\Omega}_2}{\gamma}& \pdv{\hat{\Omega}_2}{\alpha}& \pdv{\hat{\Omega}_2}{\beta}& \pdv{\hat{\Omega}_2}{\rho}& \pdv{\hat{\Omega}_2}{\sigma}& \pdv{\hat{\Omega}_2}{\lambda}&  \pdv{\hat{\Omega}_2}{n_x} & \pdv{\hat{\Omega}_2}{n_y} & \pdv{\hat{\Omega}_2}{n_z} & \pdv{\hat{\Omega}_2}{L_x}& \pdv{\hat{\Omega}_2}{L_y} & \pdv{\hat{\Omega}_2}{L_z}
 \vspace{0.2cm} \\ 
 \pdv{\hat{\Omega}_3}{\gamma}& \pdv{\hat{\Omega}_3}{\alpha}& \pdv{\hat{\Omega}_3}{\beta}& \pdv{\hat{\Omega}_3}{\rho}& \pdv{\hat{\Omega}_3}{\sigma}& \pdv{\hat{\Omega}_3}{\lambda}&  \pdv{\hat{\Omega}_3}{n_x} & \pdv{\hat{\Omega}_3}{n_y} & \pdv{\hat{\Omega}_3}{n_z} & \pdv{\hat{\Omega}_3}{L_x}& \pdv{\hat{\Omega}_3}{L_y} & \pdv{\hat{\Omega}_3}{L_z}
 \vspace{0.2cm} \\ 
 \pdv{\hat{\Omega}_4}{\gamma}& \pdv{\hat{\Omega}_4}{\alpha}& \pdv{\hat{\Omega}_4}{\beta}& \pdv{\hat{\Omega}_4}{\rho}& \pdv{\hat{\Omega}_4}{\sigma}& \pdv{\hat{\Omega}_4}{\lambda}&  \pdv{\hat{\Omega}_4}{n_x} & \pdv{\hat{\Omega}_4}{n_y} & \pdv{\hat{\Omega}_4}{n_z} & \pdv{\hat{\Omega}_4}{L_x}& \pdv{\hat{\Omega}_4}{L_y} & \pdv{\hat{\Omega}_4}{L_z}
 \end{bmatrix} . 
\end{align}
In the above matrix, 
each element represents three consecutive entries along the column, corresponding to the three components of the $\hat{\Omega}_j$ vector. Using the transformation relations in Eqs.~\ref{eqB2}, we write the matrix in Eq.~\ref{eqB5} as $$\Big[A_{12\times 6}| B_{12\times 6} \Big],$$
where 
\begin{align}
A|_{\rho =\sigma =\lambda =1} = \begin{bmatrix}
\pdv{R}{\gamma}\left(\hat{n} + M_1\frac{\vec{L}}{S}\right)& \pdv{R}{\alpha}\left(\hat{n} + M_1\frac{\vec{L}}{S}\right)& \pdv{R}{\beta}\left(\hat{n} + M_1\frac{\vec{L}}{S}\right)& \mathbf{0}& \mathbf{0}& \mathbf{0}
 \vspace{0.2cm}\\  
 \pdv{R}{\gamma}\left(T_1\hat{n} + M_2\frac{\vec{L}}{S}\right)& \pdv{R}{\alpha}\left(T_1\hat{n} + M_2\frac{\vec{L}}{S}\right)& \pdv{R}{\beta}\left(T_1\hat{n} + M_2\frac{\vec{L}}{S}\right)& RT_1\hat{n}& \mathbf{0}& \mathbf{0}
\vspace{0.2cm}\\
 \pdv{R}{\gamma}\left(T_2\hat{n} + M_3\frac{\vec{L}}{S}\right)& \pdv{R}{\alpha}\left(T_2\hat{n} + M_3\frac{\vec{L}}{S}\right)& \pdv{R}{\beta}\left(T_2\hat{n} + M_3\frac{\vec{L}}{S}\right)&\mathbf{0}& RT_2\hat{n}& \mathbf{0}
  \vspace{0.2cm}\\
 \pdv{R}{\gamma}\left(T_3\hat{n} + M_4\frac{\vec{L}}{S}\right)& \pdv{R}{\alpha}\left(T_3\hat{n} + M_4\frac{\vec{L}}{S}\right)& \pdv{R}{\beta}\left(T_3\hat{n} + M_4\frac{\vec{L}}{S}\right)& \mathbf{0}& \mathbf{0}&  RT_3\hat{n}
 \end{bmatrix}  
\end{align}
and
\begin{align}
B|_{\rho =\sigma =\lambda =1} = \begin{bmatrix}
 R\left( \mathbf{1}_{fc} + \pdv{M_1}{n_x}\frac{\vec{L}}{S}\right) & R\left( \mathbf{1}_{sc} + \pdv{M_1}{n_y}\frac{\vec{L}}{S}\right)&  R\left( \mathbf{1}_{tc} + \pdv{M_1}{n_z}\frac{\vec{L}}{S}\right)& \frac{1}{S} RM_1
\vspace{0.2cm}\\
R\left( {T_1}_{fc} + \pdv{M_2}{n_x}\frac{\vec{L}}{S}\right) &  R\left( {T_1}_{sc} + \pdv{M_2}{n_y}\frac{\vec{L}}{S}\right)&  R\left({T_1}_{tc} + \pdv{M_2}{n_z}\frac{\vec{L}}{S}\right)& \frac{1}{S} RM_2
\vspace{0.2cm}\\
 R\left( {T_2}_{fc} + \pdv{M_3}{n_x}\frac{\vec{L}}{S}\right) & R\left( {T_2}_{sc} + \pdv{M_3}{n_y}\frac{\vec{L}}{S}\right)& R\left({T_2}_{tc} + \pdv{M_3}{n_z}\frac{\vec{L}}{S}\right)& \frac{1}{S} RM_3
\vspace{0.2cm}\\
R\left( {T_3}_{fc} + \pdv{M_4}{n_x}\frac{\vec{L}}{S}\right) &  R\left( {T_3}_{sc} + \pdv{M_4}{n_y}\frac{\vec{L}}{S}\right)&  R\left({T_3}_{tc} + \pdv{M_4}{n_z}\frac{\vec{L}}{S}\right)& \frac{1}{S} RM_4
\end{bmatrix}.
\end{align}
Here $fc, sc$ and $tc$ denote the first, second and third columns of the corresponding $T$ matrix respectively. $\mathbf{0}$ is a $3\times 1$ column matrix with all entries equal to zero while $\mathbf{1}$ is a $3\times 3$ identity matrix.
At this stage, the Jacobian matrix can be written as a product of two matrices, $C$ and $D$, which are given as follows.
\begin{align}
C = 
\begin{bmatrix}[c|c|c]
R & 0 & 0\\
\hline
 0& R &  0\\
\hline
 0& 0 &  R
 \end{bmatrix},
\end{align}
a block diagonal matrix. The $3\times 3$ diagonal blocks are rotation matrices. Matrix $D$ is given by $D\equiv\Big[E_{12\times 6}| F_{12\times 6} \Big],$
where
\begin{align}
E = \begin{bmatrix}
  R^{-1}\pdv{ R}{\gamma}\left(\hat{n} + M_1\frac{\vec{L}}{S}\right)&  R^{-1}\pdv{R}{\alpha}\left(\hat{n} + M_1\frac{\vec{L}}{S}\right)&  R^{-1}\pdv{ R}{\beta}\left(\hat{n} + M_1\frac{\vec{L}}{S}\right)& \mathbf{0}& \mathbf{0}& \mathbf{0}
 \vspace{0.2cm} \\ 
  R^{-1}\pdv{ R}{\gamma}\left(T_1\hat{n} + M_2\frac{\vec{L}}{S}\right)&  R^{-1}\pdv{ R}{\alpha}\left(T_1\hat{n} + M_2\frac{\vec{L}}{S}\right)& R^{-1}\pdv{R}{\beta}\left(T_1\hat{n} + M_2\frac{\vec{L}}{S}\right)& T_1\hat{n}& \mathbf{0}& \mathbf{0}
 \vspace{0.2cm} \\ 
  R^{-1}\pdv{ R}{\gamma}\left(T_2\hat{n} + M_3\frac{\vec{L}}{S}\right)&  R^{-1}\pdv{ R}{\alpha}\left(T_2\hat{n} + M_3\frac{\vec{L}}{S}\right)& R^{-1}\pdv{ R}{\beta}\left(T_2\hat{n} + M_3\frac{\vec{L}}{S}\right)&\mathbf{0}& T_2\hat{n}& \mathbf{0}
 \vspace{0.2cm} \\ 
  R^{-1}\pdv{ R}{\gamma}\left(T_3\hat{n} + M_4\frac{\vec{L}}{S}\right)& R^{-1}\pdv{R}{\alpha}\left(T_3\hat{n} + M_4\frac{\vec{L}}{S}\right)& R^{-1}\pdv{R}{\beta}\left(T_3\hat{n} + M_4\frac{\vec{L}}{S}\right)& \mathbf{0}& \mathbf{0}& T_3\hat{n}
 \end{bmatrix} 
\end{align}
and
\begin{align}
F = \begin{bmatrix}
\left( \mathbf{1}_{fc} + \pdv{M_1}{n_x}\frac{\vec{L}}{S}\right) & \left( \mathbf{1}_{sc} + \pdv{M_1}{n_y}\frac{\vec{L}}{S}\right)& \left( \mathbf{1}_{tc} + \pdv{M_1}{n_z}\frac{\vec{L}}{S}\right)& \frac{1}{S}M_1
\vspace{0.2cm}\\
\left( {T_1}_{fc} + \pdv{M_2}{n_x}\frac{\vec{L}}{S}\right) & \left( {T_1}_{sc} + \pdv{M_2}{n_y}\frac{\vec{L}}{S}\right)& \left({T_1}_{tc} + \pdv{M_2}{n_z}\frac{\vec{L}}{S}\right)& \frac{1}{S}M_2
\vspace{0.2cm}\\
\left( {T_2}_{fc} + \pdv{M_3}{n_x}\frac{\vec{L}}{S}\right) & \left( {T_2}_{sc} + \pdv{M_3}{n_y}\frac{\vec{L}}{S}\right)& \left({T_2}_{tc} + \pdv{M_3}{n_z}\frac{\vec{L}}{S}\right)& \frac{1}{S}M_3
\vspace{0.2cm}\\
\left( {T_3}_{fc} + \pdv{M_4}{n_x}\frac{\vec{L}}{S}\right) & \left( {T_3}_{sc} + \pdv{M_4}{n_y}\frac{\vec{L}}{S}\right)& \left({T_3}_{tc} + \pdv{M_4}{n_z}\frac{\vec{L}}{S}\right)& \frac{1}{S}M_4
\end{bmatrix}.
\end{align}
The Jacobian, $\mathcal{J}_1 = Det(CD)= Det(D)$ as $Det(C) = (Det(R))^3 = 1$. The determinant of $D$ is much simpler to evaluate than that of the Jacobian matrix we started out with. Using Mathematica, $Det(D)$ can be shown to be $\varpropto\frac{1}{4\pi^2}\sin^2(\frac{\gamma}{2})\sin\alpha\, Det(M)$, to $\mathcal{O}(S^0)$. Now, we can rewrite Eq.~\ref{eq22}, apart from a constant multiplicative factor, as 
\begin{eqnarray}
\nonumber &\int\,\frac{1}{4\pi^2}\sin^2(\frac{\gamma}{2})\sin\alpha\, Det(M)\,d\vec{L}\,d\alpha\,d\beta\,d\gamma\,d\vec{n}\,\delta(n^2 -1)\\
&\sim \int\,\frac{1}{4\pi^2}\sin^2(\frac{\gamma}{2})\sin\alpha\, Det(M)\, \sin\theta\, d\theta\,d\phi\,d\vec{L}\,d\alpha\,d\beta\,d\gamma.
\end{eqnarray}
Finally the full Jacobian, $\mathcal{J}$ is proportional to $\frac{1}{4\pi^2}\sin^2(\frac{\gamma}{2})\sin\alpha\, Det(M)\, \sin\theta $.

\section{Degeneracy from the conventional quantum approach}\label{appC}
Let us count the degeneracies of the first few states using standard angular momentum addition. Adding pairs of spins, we have
\begin{align*}
S\otimes S\otimes S \otimes S&= (0\oplus 1 \oplus 2 \oplus 3 \oplus\cdots\oplus 2S)\otimes (0\oplus 1 \oplus 2 \oplus 3 \oplus\cdots\oplus 2S).
\end{align*}
If we are to obtain a net singlet, the net $S$ must be chosen to be $0$. This arises in the following cases: $(0\otimes 0), (1\otimes 1), (2\otimes 2), \cdots, (2S\otimes 2S)$. Thus, we have $2S + 1$ possible ways of having zero net spin. This is the degeneracy of the ground state. The first excited state requires the net spin to be unity. This arises from  
\begin{align*}
&(0\otimes 1),\\
(1\otimes 0), &(1\otimes 1)\; \mathrm{and}\; (1\otimes 2),\\
(2\otimes 1), &(2\otimes 2)\; \mathrm{and}\; (2\otimes 3),\\
(3\otimes 2), &(3\otimes 3)\; \mathrm{and}\; (3\otimes 4),\\
\vdots\\
(2S-1\otimes 2S-2), &(2S-1\otimes 2S-1)\; \mathrm{and}\; (2S-1\otimes 2S),\\
(2S\otimes  2S&-1)\;\mathrm{and}\;(2S\otimes 2S).
\end{align*}
This accounts for $3(2S -1) + 3 = 3 (2S + 1) - 3$ possibilities. In addition, each state with net spin unity has a threefold degeneracy corresponding to three choices of the $S_z$ quantum number. The full degeneracy of the first excited state is given by $3^2 (2S + 1) - 9$.

The second excited state must have net spin 2. This arises from 
\begin{align*}
&(0\otimes 2),\\
(1\otimes 1), &(1\otimes 2)\; \mbox{and}\; (1\otimes 3),\\
(2\otimes 0), (2\otimes 1), &(2\otimes 2), (2\otimes 3)\;\mbox{and}\; (2\otimes 4),\\
(3\otimes 1), (3\otimes 2),&(3\otimes 3), (3\otimes 4)\;\mbox{and}\; (3\otimes 5),\\
\vdots\\
(2S - 2\otimes 2S-4), (2S-2\otimes 2S-3), &(2S-2\otimes 2S-2), (2S-2\otimes 2S-1)\;\mbox{and}\; (2S-2\otimes 2S),\\
(2S-1\otimes 2S-3), (2S-1\otimes &2S-2),(2S-1\otimes 2S-1)\; \mbox{and}\; (2S-1\otimes 2S),\\
(2S\otimes  2S-2),(2S&\otimes  2S-1)\;\mbox{and}\;(2S\otimes 2S).
\end{align*}
This amounts to $5(2S -3) + 1 + 3+4 + 3 = 5(2S + 1) -9$ possibilities. Each of these states has a five-fold degeneracy. On the whole, the second excited state has degeneracy $5^2(2S +1) -45$. The degeneracy for higher excited states can be enumerated in a similar fashion. 

\section{Implicit function theorem on the quadrumer ground state space }\label{implicit.func.proof}
Let us denote the positions of four spins in our problem as $\vec{S}_1\equiv (x_1, y_1, z_1)$,  $\vec{S}_2\equiv (x_2, y_2, z_2)$, $\vec{S}_3\equiv (x_3, y_3, z_3)$ and $\vec{S}_4\equiv (x_4, y_4, z_4)$. Let $\mathbf{x}:=(x_1, y_1, z_1, x_2, y_2, z_2, \cdots, x_4, y_4, z_4) \in\mathbb{R}^{12}$ and $\mathbf{f}$ be a map from $\mathbb{R}^{12}\mapsto\mathbb{R}^7$ given by
\begin{equation} 
\mathbf{f}:= (f_1, f_2, f_3, f_4, f_5, f_6, f_7),
\end{equation}
where
\begin{align}
f_1&:= x_1 + x_2 + x_3 + x_4,\nonumber\\
f_2&:= y_1 + y_2 + y_3 + y_4,\nonumber\\
f_3&:= z_1 + z_2 + z_3 + z_4,\nonumber\\
f_4&:= x_1^2 + y_1^2 + z_1^2, \nonumber\\
f_5&:= x_2^2 + y_2^2 + z_2^2, \nonumber\\
f_6&:= x_3^2 + y_3^2 + z_3^2, \nonumber\\
f_7&:= x_4^2 + y_4^2 + z_4^2.
\label{eq.constraints}
\end{align}
The quantities $f_{1-3}$ represent the vector sum of the four spins. Likewise, $f_{4-7}$ represent the magnitudes of the four spins.
Using the implicit function theorem, we wish to examine if $\mathbf{f}^{-1}(0, 0, 0, 1, 1, 1, 1)$ is a manifold. We are interested in the inverse image of a specific point $(0, 0, 0, 1, 1, 1, 1)$ as it corresponds to the ground state criterion in our quadrumer problem. In other words, we wish  to know if our ground state space forms a manifold in the precise mathematical sense.

The implicit function theorem requires the construction of a `Jacobian matrix'. If this matrix has full rank (for an $m\times n$ matrix, it is $m (n)$ if $m\leq (\geq) n$) at every point in the ground state space, then the space is a manifold. In particular, it is a manifold of dimension $D-C$, where $D$ is the dimension of the embedding space (12 in our case) and $C$ is the number of constraints (7 in our case). Here, the Jacobian matrix is given by,
\begin{align*}
J(\mathbf{f})_{7\times 12}=\begin{bmatrix}
    \pdv{f_{1}}{x_1}& \pdv{f_{1}}{y_1}& \pdv{f_{1}}{z_1}& \pdv{f_{1}}{x_2}& \pdv{f_{1}}{y_2}& \pdv{f_{1}}{z_2}& \cdots & \pdv{f_{1}}{z_4}
\vspace{0.2cm} \\    
\pdv{f_{2}}{x_1}& \pdv{f_{2}}{y_1}& \pdv{f_{2}}{z_1}& \pdv{f_{2}}{x_2}& \pdv{f_{2}}{y_2}& \pdv{f_{2}}{z_2}& \cdots & \pdv{f_{2}}{z_4}
\vspace{0.2cm} \\    
\pdv{f_{3}}{x_1}& \pdv{f_{3}}{y_1}& \pdv{f_{3}}{z_1}& \pdv{f_{3}}{x_2}& \pdv{f_{3}}{y_2}& \pdv{f_{3}}{z_2}& \cdots & \pdv{f_{3}}{z_4}
 \vspace{0.2cm} \\    
 \pdv{f_{4}}{x_1}& \pdv{f_{4}}{y_1}& \pdv{f_{4}}{z_1}& \pdv{f_{4}}{x_2}& \pdv{f_{4}}{y_2}& \pdv{f_{4}}{z_2}& \cdots & \pdv{f_{4}}{z_4}    
 \vspace{0.2cm} \\    
 \pdv{f_{5}}{x_1}& \pdv{f_{5}}{y_1}& \pdv{f_{5}}{z_1}& \pdv{f_{5}}{x_2}& \pdv{f_{5}}{y_2}& \pdv{f_{5}}{z_2}& \cdots & \pdv{f_{5}}{z_4}
 \vspace{0.2cm} \\    
 \pdv{f_{6}}{x_1}& \pdv{f_{6}}{y_1}& \pdv{f_{6}}{z_1}& \pdv{f_{6}}{x_2}& \pdv{f_{6}}{y_2}& \pdv{f_{6}}{z_2}& \cdots & \pdv{f_{6}}{z_4}
 \vspace{0.2cm} \\    
 \pdv{f_{7}}{x_1}& \pdv{f_{7}}{y_1}& \pdv{f_{7}}{z_1}& \pdv{f_{7}}{x_2}& \pdv{f_{7}}{y_2}& \pdv{f_{7}}{z_2}& \cdots & \pdv{f_{7}}{z_4}. 
\end{bmatrix} = 
\begin{bmatrix}
1 & 0 & 0 & 1 & 0 & 0 & 1 & 0 & 0 & 1 & 0 & 0\\
0 & 1 & 0 & 0 & 1 & 0 & 0 & 1 & 0 & 0 & 1 & 0\\
0 & 0 & 1 & 0 & 0 & 1 & 0 & 0 & 1 & 0 & 0 & 1\\
2x_1 & 2y_1 & 2z_1& 0 & 0 & 0 & 0& 0 & 0 & 0 & 0 & 0\\
0 & 0 & 0 & 2x_2 & 2y_2 & 2z_2& 0& 0 & 0 & 0 & 0 & 0\\
0& 0 & 0 & 0 & 0 & 0 &  2x_3 & 2y_3 & 2z_3& 0 & 0 & 0\\
0& 0 & 0 & 0 & 0 & 0 & 0 & 0 & 0 & 2x_4 & 2y_4 & 2z_4
\end{bmatrix}.
\end{align*} 
To determine the rank of the matrix, we determine the number of independent rows here. We denote the rows as twelve-component vectors $r_i$, with $i=1,\ldots,7$.
We consider the set of linear equations 
\begin{equation}\label{eq.lindep}
\sum_{i=1}^7 \lambda_i r_i = 0.
\end{equation}
If this possesses a non-trivial solution (i.e., with at least one $\lambda$ being non-zero), the rows are not independent and the Jacobian matrix does not have full rank.
Equation~\ref{eq.lindep} reduces to the following twelve equations.
\begin{eqnarray}\label{eq.lindep1}
\begin{matrix}
\lambda_1 + 2x_1\lambda_4 = 0, & \lambda_1 + 2x_2\lambda_5 = 0, & \lambda_1 + 2x_3\lambda_6 = 0, & \lambda_1 + 2x_4\lambda_7 = 0,\\
\lambda_2 + 2y_1\lambda_4 = 0, & \lambda_2 + 2y_2\lambda_5 = 0, & \lambda_2 + 2y_3\lambda_6 = 0, & \lambda_2 + 2y_4\lambda_7 = 0,\\
\lambda_3 + 2z_1\lambda_4 = 0, & \lambda_3 + 2z_2\lambda_5 = 0,& \lambda_3 + 2z_3\lambda_6 = 0, & \lambda_3 + 2z_4\lambda_7 = 0.
\end{matrix}
\end{eqnarray}
We consider these equations at points on the ground state space, i.e., at points which satisfy
\begin{eqnarray}
\sum_{i=1}^4x_i = 0, \phantom{ab}\sum_{i=1}^4y_i = 0,\phantom{ab} \sum_{i=1}^4z_i = 0\label{eq.sumzero}\\
x_j^2 + y_j^2 + z_j^2 = 1,\,\,j= 1, 2, 3,4.\label{eq.norm}
\end{eqnarray}
We first consider collinear ground states and show that a non-trivial solution for Eq.~\ref{eq.lindep} exists. In this case, we show that the rank of the Jacobian matrix is six. We next consider non-collinear states and show that we only have a trivial solution.

\noindent\underline{Collinear states}:
Without loss of generality, we consider a collinear state with $\vec{S}_1=\vec{S}_2=-\vec{S}_3=-\vec{S}_4$, i.e.,
\begin{eqnarray}
\nonumber &x_1 = x_2= -x_3 = -x_4\\
\nonumber &y_1 = y_2= -y_3 = -y_4\\
&z_1 = z_2= -z_3 = -z_4.
\label{eq.coll}
\end{eqnarray}
The following proof can be easily extended to other cases as well.
Using Eqs.~ \ref{eq.coll} and \ref{eq.lindep1} (note that $x_j$, $y_j$ and $z_j$ cannot all be zero), we obtain
\begin{equation}\label{eq.lambda}
\lambda_4= \lambda_5 = -\lambda_6 = -\lambda_7.
\end{equation}
If $\lambda_7$ is zero, we immediately find that all $\lambda$'s vanish, leading to a trivial solution. Assuming a non zero value for $\lambda_7$, it is easy to find suitable values of all other $\lambda$'s using Eqs.~\ref{eq.lambda} and \ref{eq.lindep1}. Thus, a non-trivial 
solution exists and the seven rows are not linearly independent. 
In other words, the rank of the matrix is less than seven.
In this light, we check if the first six rows of the matrix are linearly independent.
Taking Eq.~\ref{eq.lindep1} with seventh row excluded, we obtain the following set of equations,
\begin{eqnarray}\label{eq.lindepcoll}
\begin{matrix}
\lambda_1 + 2x_1\lambda_4 = 0, & \lambda_1 + 2x_2\lambda_5 = 0, & \lambda_1 + 2x_3\lambda_6 = 0, & \lambda_1  = 0,\\
\lambda_2 + 2y_1\lambda_4 = 0, & \lambda_2 + 2y_2\lambda_5 = 0, & \lambda_2 + 2y_3\lambda_6 = 0, & \lambda_2 = 0,\\
\lambda_3 + 2z_1\lambda_4 = 0, & \lambda_3 + 2z_2\lambda_5 = 0,& \lambda_3 + 2z_3\lambda_6 = 0, & \lambda_3 = 0.
\end{matrix}
\end{eqnarray}
It is clear from Eq.~\ref{eq.lindepcoll} that all $\lambda$'s must vanish. Therefore, we have six linearly independent rows in the Jacobian matrix. We conclude that the rank of the Jacobian matrix is 6 for any collinear state.\

\noindent\underline{Non-collinear states}:
We have a strong constraint on the $\lambda$'s, viz., $\lambda_4$, $\lambda_5$, $\lambda_6$ and $\lambda_7$ must be non zero for a non-trivial solution to exist. If any one of them is zero, Eq.~\ref{eq.lindep1} immediately forces all $\lambda$'s to be zero. We now rewrite Eqs.~\ref{eq.sumzero} using Eq.~\ref{eq.lindep1} to obtain
\begin{eqnarray}
\nonumber \lambda_1\left(\frac{1}{\lambda_4} + \frac{1}{\lambda_5} + \frac{1}{\lambda_6} + \frac{1}{\lambda_7}\right) = 0,\\
\nonumber \lambda_2\left(\frac{1}{\lambda_4} + \frac{1}{\lambda_5} + \frac{1}{\lambda_6} + \frac{1}{\lambda_7}\right) = 0,\\ 
\lambda_3\left(\frac{1}{\lambda_4} + \frac{1}{\lambda_5} + \frac{1}{\lambda_6} + \frac{1}{\lambda_7}\right) = 0.
\end{eqnarray}
We argue that $\left(\frac{1}{\lambda_4} + \frac{1}{\lambda_5} + \frac{1}{\lambda_6} + \frac{1}{\lambda_7}\right)$ must vanish. Otherwise, we will have $\lambda_1 = \lambda_2 = \lambda_3 = 0$. This will in turn force $\lambda_{4,\ldots ,7}$ to also vanish due to Eq.~\ref{eq.lindep1}, leading to a trivial solution. As a result, a non-trivial solution requires
\begin{equation}\label{eq.lambda1}
\left(\frac{1}{\lambda_4} + \frac{1}{\lambda_5} + \frac{1}{\lambda_6} + \frac{1}{\lambda_7}\right) = 0.
\end{equation}
From eq.~\ref{eq.lindep1} and eq.~\ref{eq.norm}, we have $\lambda_1^2 + \lambda_2^2 + \lambda_3^2 = 2\lambda_4^2 = 2\lambda_5^2 = 2\lambda_6^2 = 2\lambda_7^2$. We find that $\lambda_{4,\ldots, 7}$ have the same magnitude. In order to satisfy Eq.~\ref{eq.lambda1}, we must have
\begin{equation}
\lambda_4=\lambda_5=-\lambda_6 = -\lambda_7,
\end{equation} 
or equivalently, we could pick two others to be negative. Using Eq.~\ref{eq.lindep1}, this relation reduces to Eq.~\ref{eq.coll} -- a collinearity condition for the ground state. We have shown that a non-trivial solution for $\lambda$'s exists only at collinear ground states. 
Therefore, at all non-collinear ground states, the Jacobian matrix has full rank. This guarantees that the ground state space, \textit{after excluding collinear states}, is a five-dimensional manifold. 
\section{Soft fluctuations around a collinear state }
\label{app.collinearsoft}
A general state of the system is represented by a twelve dimensional vector $(\vec{S}_1, \vec{S}_2, \vec{S}_3, \vec{S}_4)$. A collinear choice for the ground state is given by $\vec{\Sigma} = (0,0,1,0,0,1,0,0,-1,0,0,-1)$. This state is shown pictorially in Fig.~\ref{fig3}, alongwith six `soft' deformations which do not cost energy. These six fluctuations about this state are explicitly given by  
\begin{align}
\vec{\sigma}_1 = (0,-1,0,0,-1,0,0,1,0,0,1,0),\nonumber\\
\vec{\sigma}_2 = (1,0,0,1,0,0,-1,0,0,-1,0,0), \nonumber \\
\vec{\sigma}_3 = (1,0,0,-1,0,0,-1,0,0,1,0,0), \nonumber\\
\vec{\sigma}_4 = (1,0,0,-1,0,0,1,0,0,-1,0,0), \nonumber\\
\vec{\sigma}_5 = (0,-1,0,0,1,0,0,1,0,0,-1,0), \nonumber\\
\vec{\sigma}_6 = (0,-1,0,0,1,0,0,-1,0,0,1,0).
\end{align}
These fluctuations modes, in this same order, are shown in Figs.~\ref{fig3}(b)-(g). They are easily seen to independent, as they are orthogonal in the twelve-dimensional embedding space. 
To see their soft character, we consider a point $ \vec{P}= \vec{\Sigma} + \sum _{i=1}^6 \delta_i \vec{\sigma}_i$, a deviation from the collinear state $\vec{\Sigma}$. The $\delta_i$'s are the amplitudes of small deviations along these six directions. 
It is easy to see that this new point $\vec{P}$ satisfies the seven ground state constraints given in Eqs.~\ref{eq.constraints} to linear order in $\delta_i$'s.
Thus, $\vec{P}$ also lies on the ground state manifold for infinitesimal $\delta_i$'s. 
This demonstrates that $\vec{\sigma} $'s form a six dimensional tangent space around $\vec{\Sigma} $. Apart from these modes, the system can have six more independent fluctuations as the space is twelve-dimensional. Two of these represent `hard' modes as they lead to a non-zero total spin and thereby incur an energy cost. The remaining four are unphysical as they violate fixed spin-length constraints.
 \end{widetext}
\section{Spin wave theory about a collinear state}\label{spinwave}
The path integral for the quadrumer presented in Eq.~\ref{eq.partitionfunction} is valid at all ground states. However, the identification with a rigid rotor fails at collinear states. Another important difference emerges at the next step when integrating out hard modes. The number of hard modes is different at collinear states. This prevents a uniform low energy description (in terms of soft modes alone) encompassing both collinear and non-collinear states. Here, we describe a `spin-wave' description of the path integral assuming that the system always remains in the vicinity of a collinear state. 

We consider a N\'eel-like configuration with ordered moments along the X axis. We reexpress the rotation matrix $R = e^W$, where $W$ lives in the Lie algebra space of $SO(3)$. We write $W = i\sum_j\pi_j T_j$, where $T_j$'s are the generators of rotation about coordinate axes. The rotation amplitudes $\pi_j$ are assumed to remain small, so that we are always in the vicinity of the reference state.
\[
T_x=
  \begin{bmatrix}
    0 & 0 & 0\\
    0 & 0 & i\\
    0 & -i & 0 
  \end{bmatrix},
  T_y=
  \begin{bmatrix}
    0 & 0 & -i\\
    0 & 0 & 0\\
    i & 0 & 0 
  \end{bmatrix},
  T_z=
  \begin{bmatrix}
    0 & i & 0\\
    -i & 0 & 0\\
    0 & 0 & 0 
  \end{bmatrix}.
\]
$\pi_x, \pi_y$ and $\pi_z $ are rotations about X, Y and Z axes respectively. Our reference state is described by $(\theta = \frac{\pi}{2}, \phi = 0)$. To second order in fluctuations $(\delta\theta = \theta - \frac{\pi}{2}, \delta\phi = \phi - 0, \pi_x, \pi_y, \pi_z, L_x, L_y, L_z)$, we get the action
\begin{equation}
\mathcal{S} = \int_0^\beta d\tau\left(iS\,\delta\theta\,\delta\dot{\phi} - 4 (L_y\dot{\pi}_y + L_z\dot{\pi}_z) + 16 J(L_y^2 + L_z^2)\right).
\end{equation}
The Jacobian, to second order in fluctuations, comes out to be proportional to $16(\delta\theta)^2 + (\delta\phi)^2$. The partition function in terms of the fluctuations is
\begin{eqnarray}\label{eq33}
\mathcal{Z} = \int \left[\prod_\tau \left\{16(\delta\theta)_\tau^2 + (\delta\phi)_\tau^2\right\}\,d{\pi_x}_\tau\,d{\pi_y}_\tau\,d{\pi_z}_\tau\right.\times\nonumber\\
\left.d(\delta\theta)_\tau\,d(\delta\phi)_\tau\,d\vec{L}_\tau\right] e^{-\mathcal{S}}.
\end{eqnarray} 
The dependence on $(L_i, \pi_i)$ variables has an identifiable form. It is the path integral of a `rigid rod' with only two rotation degrees of freedom. The parameter $L_x$ does not appear in the action. This is in line with our parametrization of collinear states, with one component of $\vec{L}$ becoming redundant. Similarly, $\pi_x$ does not appear as it corresponds to a trivial rotation about the axis of collinear order. The hard modes are $L_y$ and $L_z$, which can now be integrated out. This form of the path integral is drastically different from that obtained around any non-collinear state, wherein all three components of $\vec{L}$ represent hard modes. This brings out the non-manifold character of the ground state space.
\bibliographystyle{apsrev4-1} 
\bibliography{quadrumer}
\end{document}